\documentclass[12pt]{article}
\usepackage{cite}
\usepackage[T1]{fontenc}
\usepackage[utf8]{inputenc}
\usepackage{bigints}
\usepackage[english]{babel}
\usepackage{microtype}

\usepackage{wrapfig}
\usepackage{graphicx}
\usepackage[toc,page]{appendix}
\usepackage[dvipsnames]{xcolor}
\usepackage{float}
\usepackage{tikz}
\usepackage[compat=1.1.0]{tikz-feynman}
\tikzfeynmanset{
  warn luatex=false,
  /graph drawing/level distance/.initial=,
  /graph drawing/sibling distance/.initial=,
  /tikzfeynman/warn luatex=false
}
\usepackage{adjustbox}
\usepackage{amsmath}
\usepackage{amsfonts}
\usepackage{amssymb}
\usepackage{amstext}
\usepackage{slashed}
\usepackage{braket}
\usepackage{caption}
\usepackage{subcaption}
\usepackage{layouts}
\usepackage[normalem]{ulem}
\usepackage{physics}

\usepackage[bookmarks, breaklinks, colorlinks,urlcolor=black, citecolor=red, linkcolor=blue]{hyperref}
\usepackage{booktabs}
\usepackage{lscape}
\usepackage{adjustbox}
\usepackage{rotating}
\usepackage{array}

\def\dsp{\displaystyle}
\def\bea{\begin{align}}
\def\eea{\end{align}} 
\def\be{\begin{equation}}
\def\ee{\end{equation}} 
\def\nn{\nonumber}
\def\Re{\text{Re}}
\def\Im{\text{Im}}
\def\tev{\ensuremath{\mathrm{Te\kern -0.1em V}}}
\def\gev{\ensuremath{\mathrm{Ge\kern -0.1em V}}}
\def\mev{\ensuremath{\mathrm{Me\kern -0.1em V}}}
\def\mdm{\ensuremath{m_\text{DM}}}
\def\gdm{\ensuremath{g_\text{DM}}}

\begin{document}
	
\begin{flushright}
SI-HEP-2026-04
\end{flushright}

\vskip 2cm	
	
	\begin{center}
		
		{\Large\bf Connecting Flavor and Baryon Asymmetry via \\[2pt] Leptogenesis in Effective Froggatt-Nielsen Theory } \\[8mm]
		 {Cheshta Batra$^a$\,\footnote{Email: cheshtacheshta@iitgn.ac.in }, Rusa Mandal$^a$\,\footnote{Email: rusa.mandal@iitgn.ac.in}, Kunal Rawat$^a$\,\footnote{Email: kunal.rawat@iitgn.ac.in } and Tom Tong$^{b,c}$\,\footnote{Email: ttong@aip.de}}
    \vskip 5pt
    
   {\small\em $^a$Indian Institute of Technology Gandhinagar, Department of Physics, \\ Gujarat 382355, India}

     {\small\em $^b$Center for Particle Physics Siegen, Universit\"at Siegen, \\ 57068 Siegen, Germany}

     {\small\em $^c$Leibniz Institute for Astrophysics Potsdam, 14482 Potsdam, Germany}

	\end{center}

\begin{abstract}

We investigate the hierarchical flavor structure of the Standard Model (SM) in a Froggatt-Nielsen (FN) framework, where the spontaneous breaking of a $U(1)_{\rm FN}$ symmetry by a complex flavon field generates fermion masses and mixing patterns through higher-dimensional operators. Extending the setup with three right-handed neutrinos (RHNs), light neutrino masses arise via the Type-I seesaw mechanism. Allowing complex FN coefficients enables a consistent description of the CKM and PMNS matrices while inducing CP-violating signatures in meson decays. Building on our previous work, where the lightest RHN acts as a viable dark matter (DM) candidate produced through freeze-in or freeze-out mechanisms, we investigate the origin of the baryon asymmetry of the Universe. The heavier RHNs generate a lepton asymmetry through out-of-equilibrium decays and scatterings, including both SM channels and additional flavon-induced processes in which the flavon appears as an initial-state particle. We compute the corresponding one-loop CP asymmetries and incorporate these effects in the Boltzmann equations. We show that although freeze-in and freeze-out DM production occur in two qualitatively distinct regions of the FN symmetry-breaking scale $v_\phi$, successful thermal leptogenesis can be achieved in both regimes. In the large-$v_\phi$ (freeze-in-compatible) region, the results approach the standard leptogenesis limit, while in the freeze-out-compatible region the lower value of $v_\phi$ implies lighter RHNs, requiring resonant enhancement. This tightly constrained framework, in which $v_\phi$ simultaneously controls RHN masses and the interaction strengths of the flavon and DM sectors, provides a predictive and unified description of flavor hierarchies, neutrino masses, CP violation, DM, and baryogenesis within a single effective theory.

\end{abstract}

\newpage
{\hypersetup{linkcolor=black}\tableofcontents}
\section{Introduction}

Within the Standard Model (SM), fermion Yukawa couplings span several orders of magnitude, with no underlying explanation for their observed structure. The Froggatt-Nielsen (FN) mechanism provides an elegant and economical framework to address this puzzle by invoking an additional abelian flavor symmetry under which SM fermions carry generation-dependent charges~\cite{Froggatt:1978nt}. The spontaneous breaking of this symmetry, mediated by a scalar field (known as `flavon'), naturally generates hierarchical Yukawa couplings through higher-dimensional operators suppressed by a large flavor scale. Consequently, fermion masses and mixing angles emerge as powers of a small symmetry-breaking parameter, offering a unified and predictive explanation for the observed flavor hierarchies.

While no direct signal of physics beyond the SM has been observed so far, neutrino oscillation data provide unambiguous evidence for new physics~\cite{Super-Kamiokande:1998kpq,SNO:2002tuh}. The introduction of right-handed neutrinos (RHNs) within the FN framework not only enables a natural realization of the Type-I seesaw mechanism
but also allows the FN symmetry to control the structure of the neutrino mass matrix. Beyond the flavor puzzle and neutrino physics, cosmological observations demand explanations for two additional phenomena absent in the SM: the nature of Dark Matter (DM) and the origin of the baryon asymmetry of the Universe~\cite{Planck:2018vyg}.

In this work, we extend the conventional FN setup in two essential directions: first, by introducing three RHNs, and second, by allowing the effective FN coefficients to be complex. 
The inclusion of RHNs facilitates the generation of light neutrino masses via the seesaw mechanism~\cite{Ema:2016ops}, while the presence of irreducible complex FN coefficients plays a central role in reproducing the observed CP-violating phases in both the quark and lepton sectors. As a result, the Cabibbo-Kobayashi-Maskawa (CKM)~\cite{Cabibbo:1963yz,Kobayashi:1973fv} and Pontecorvo-Maki-Nakagawa-Sakata (PMNS)~\cite{Pontecorvo:1967fh,Maki:1962mu} matrices are accurately described within the same framework. These complex couplings also give rise to nontrivial flavor signatures, particularly in the quark sector, leading to testable effects in flavor observables such as rare and CP-violating $B$-meson decays.

In our previous work~\cite{Mandal:2023jnv}, we demonstrated that a minimal FN extension of the SM with three RHNs can successfully account for DM, with the flavon acting as a portal between the dark sector and the SM. In that framework, the lightest RHN emerges as a viable DM candidate, with the observed relic abundance achieved through both freeze-in and freeze-out mechanisms, while remaining consistent with all relevant phenomenological and experimental constraints.
The present work builds on this setup by addressing the origin of the baryon asymmetry. While the lightest RHN remains the DM candidate and is effectively decoupled from thermal leptogenesis dynamics, the two heavier RHNs can naturally generate the observed baryon asymmetry through standard thermal leptogenesis~\cite{Fukugita:1986hr}. The generation of the lepton asymmetry is driven not only by standard CP-violating decays but also receives contributions from an additional $2\leftrightarrow2$ scattering process induced by complex FN coefficients involving flavon in the initial state. These new CP-violating sources enhance the lepton asymmetry generated in out-of-equilibrium RHN decays, which is then converted into baryon asymmetry through electroweak sphaleron processes~\cite{Kuzmin:1985mm}. We show that successful leptogenesis can be achieved while remaining consistent with low-energy flavor observables, fermion mass hierarchies, neutrino oscillation data, and the DM phenomenology established in our earlier work. In particular, we analyze two parametrically distinct regimes of the FN symmetry-breaking scale $v_\phi$, corresponding to freeze-in and freeze-out DM production. In the freeze-in regime, $v_\phi \sim 10^{8}\,\text{GeV}$ and flavon-mediated interactions are highly suppressed, whereas in the freeze-out regime, $v_\phi \sim \mathcal{O}(1)\,\text{TeV}$ and these interactions are significantly stronger. Despite this hierarchy and the resulting differences in RHN dynamics, efficient thermal leptogenesis can be realized in both regions.

The remainder of this paper is organized as follows. In Sec.~\ref{sec:model}, we introduce the FN framework, including the field content, charge assignments, and effective interactions relevant for fermion masses and mixings. Section~\ref{sec:constraint} presents the experimental constraints, with emphasis on $B_{d,s}$-meson and kaon mixing observables. In Sec.~\ref{sec:boltzmann}, we derive the CP asymmetries and Boltzmann equations relevant for thermal leptogenesis. Numerical results are presented in Sec.~\ref{sec:results}, where we analyze parameter regions compatible with freeze-in and freeze-out DM scenarios (Secs.~\ref{sec:freeze-in} and~\ref{sec:freeze-out}). Finally, Sec.~\ref{sec:summary} summarizes our conclusions, while technical details are collected in the appendices.

\section{Froggatt-Nielsen framework}
\label{sec:model}

At energies below the FN scale, flavor dynamics are encoded in effective operators involving SM fields, the flavon, and the RHNs. The corresponding interaction Lagrangian for the flavon field $\phi$, the SM fermions, and three SM-gauge-singlet RHNs $N_R^i$ can be written as~\cite{Ema:2016ops,Mandal:2023jnv}:
\begin{align}
\label{eq:Lag}
	-\mathcal{L}_\text{int} &=c^{i j}_{d}\left(\frac{\phi}{M}\right)^{n^{i j}_{d}} \bar{Q}^{i} H d_{R}^j+c^{i j}_{u}\left(\frac{\phi}{M}\right)^{n^{i j}_{u}} \bar{Q}^{i}\;i \sigma_2H^* u_{R }^j +c^{i j}_{e}\left(\frac{\phi}{M}\right)^{n^{i j}_{e}} \bar{L}^{i} H e_{R }^j \nn \\
	&+c^{ij}_{\nu}\left(\frac{\phi}{M}\right)^{n^{i j}_{\nu}} \bar{L}^{i}\, i \sigma_2 H^* N_{R }^j +\frac{1}{2} c^{ij}_{N}\left(\frac{\phi}{M}\right)^{n^{ij}_{N}} M \overline{N_{R}^{c\,i}} N_{R}^j \ + \ \text{h.c.} \,.
\end{align}
We introduce an additional abelian $U(1)_{\rm FN}$ symmetry under which SM fermions carry generation-dependent charges. The corresponding charge differences are defined as
\begin{equation}
\begin{aligned}
	&n^{i j}_{u}\equiv q_{Q_{i}}-q_{u_{j}},~~n^{i j}_{d} \equiv q_{Q_{i}}-q_{d_{j}},~~ n^{i j}_{e} \equiv q_{L_{i}}-q_{e_{j}}, \\[1ex]
	& \quad \quad~~ n^{i k}_{\nu} \equiv q_{L_{i}}-q_{N_k},~~n^{i j}_{N} \equiv -q_{N_i}-q_{N_j} \,.
\label{eq:n-to-q}
\end{aligned}
\end{equation}
All exponents satisfy $n_f^{ij} \geq 0$, implying $q_{N_i} \leq 0$, while the coefficients $c_x^{ij}$ are taken to be $\mathcal{O}(1)$ for $i,j=1,2,3$.

Spontaneous breaking of the FN symmetry occurs when $\phi$ acquires a vacuum expectation value (vev),
\begin{align}
\label{eq:vevs}
\phi=v_{\phi}+\frac{1}{\sqrt{2}}(s+i a)\,,
\end{align}
whose effects are transmitted to SM fermions through higher-dimensional operators suppressed by powers of $\epsilon \equiv \langle \phi \rangle / M$. Here, $M$ denotes the flavor-dynamics scale, identified with the mass scale of heavy vector-like FN fields that have been integrated out. After electroweak symmetry breaking, with $H^T = \left(0, v_{\rm EW} + h/\sqrt{2}\right)$, the quark and charged-lepton mass matrices are
\be
\label{eq:massmat}
m_f^{ij}= c_f^{ij} \epsilon^{n_f^{ij}} v_\text{EW}~~\text{with}~~f=u,d,e\,.
\ee
Using the following unitary rotations from flavor to mass eigenstates,
\begin{align}
f_L^i \to U_L^{ij} f_L^j, \qquad f_R^i \to U_R^{ij} f_R^j\,,
\end{align}
the mass matrices are diagonalized. The leading dimension-four interaction terms for the scalar and pseudoscalar fields are then given by
\begin{align}
\label{eq:gaf}
-\mathcal{L}_\text{scalar} = \sum_{f=u,d,e}
\bigg[& m_i^f \left(1 + \frac{h}{\sqrt{2}v_\text{EW}} \right) \bar{f}^i f^i  \nn \\
&+  i a \left( (g_+^f)_{ij} \bar{f}^i \gamma_5 f^j + (g_-^f)_{ij} \bar{f}^i f^j  \right) \nn \\
&+ s \left( (g_+^f)_{ij} \bar{f}^i  f^j + (g_-^{f})_{ij} \bar{f}^i \gamma_5 f^j  \right) \bigg]\,,
\end{align}
where the couplings are~\cite{Ema:2016ops}
\begin{align}
\label{eq:g+}
(g_+^f)_{ij} \ &= \ \frac{1}{2 \sqrt{2}} \left( U_L^{f \dagger} \hat{q}_Q U_L^f -  U_R^{f \dagger} \hat{q}_f U_R^f  \right)_{ij} \frac{m_j^f+ m_i^f}{v_\phi}\,, \\
\label{eq:g-}
(g_-^f)_{ij} \ &= \ \frac{1}{2 \sqrt{2} } \left( U_L^{f \dagger} \hat{q}_Q U_L^f +  U_R^{f \dagger} \hat{q}_f U_R^f  \right)_{ij} \frac{m_j^f- m_i^f}{v_\phi}\,.
\end{align}
Here, $ (\hat{q}_X)_{ij}= q_{X_i} \delta_{ij}$ are the diagonal matrices of the FN charges. Note that because of the presence of generation-dependent FN charges, the flavon couplings cannot be diagonalized simultaneously with the mass matrices. As a result, flavor-changing interactions are generated.
The fermion masses and mixing pattern depend sensitively on the $U(1)_{\rm FN}$ charge assignments, in particular on charge differences. Adopting the standard choice for $\epsilon$ as the Cabibbo angle,
\begin{align}
\label{eq:eps_cabibo}
\epsilon = \frac{v_\phi}{M} \approx 0.23 \,,
\end{align}
the following charge differences reproduce the observed mass spectrum:
\begin{align}
	&n^{ij}_u=\begin{pmatrix}
		8 & 4 & 3 \\
		7 & 3 & 2 \\
		5 & 1 & 0 
	\end{pmatrix},~~
	n^{ij}_d=\begin{pmatrix}
		7 & 6 & 6 \\
		6 & 5 & 5 \\
		4 & 3 & 3 
	\end{pmatrix},~~
	n^{ij}_e=\begin{pmatrix}
		9 & 6 & 4 \\
		8 & 5 & 3 \\
		8 & 5 & 3 
	\end{pmatrix} \,.
	\label{eq:nFN}
\end{align}
Although the individual FN charges $q_X$ are not unique, their implications for the quark and lepton sectors (including the RHNs) are tightly constrained. Before discussing those details, we highlight a few key features of the model.

First, the scalar potential for the complex flavon field $\phi$ is discussed in Ref.~\cite{Mandal:2023jnv}. As $\phi$ is responsible for the spontaneous breaking of $U(1)_{\rm FN}$, the associated pseudoscalar $a$ (see Eq.~\eqref{eq:vevs}) is a Goldstone mode and is massless at this stage. A soft $U(1)_{\rm FN}$-breaking term can generate a finite mass $m_a$. The scalar component $s$ acquires a mass of order $m_s \simeq \sqrt{2}\,v_\phi$, leading to a hierarchical spectrum with $m_a \ll m_s$. As noted in Ref.~\cite{Mandal:2023jnv}, the heavy scalar $s$ decouples from low-energy phenomenology, so in the present analysis the pseudoscalar mass $m_a$ remains the only relevant free parameter in the flavon sector.

Second, as extensively explored in Ref.~\cite{Mandal:2023jnv}, the lightest of the three RHNs, $N_1$, serves as the DM candidate. Its stability is ensured by imposing an additional $Z_2$ symmetry under which only $N_1$ is odd. This symmetry forbids the couplings $c^{i1}_{\nu}=0$ and $c^{1k}_{N} = c^{k1}_{N} = 0$ for $k=2,3$, thereby preventing $N_1$ from mixing with the other RHNs and from decaying into SM states.

We now return to the discussion of the individual FN charge assignments for the fermions in our model. The quark sector is relatively straightforward: the observed quark masses are obtained by diagonalizing the mass matrices in Eq.~\eqref{eq:massmat}, while the CKM matrix is determined from the corresponding rotation matrices, given by
\be
V_{\rm CKM}=U_L^{u\dagger} U_L^d\,.
\ee
A possible assignment is
\begin{align}
\label{eq:FNsol}
\begin{gathered}
\left(\begin{array}{ccc}
q_{Q_{1}} & q_{Q_{2}} & q_{Q_{3}} \\
q_{u} & q_{c} & q_{t} \\
q_{d} & q_{s} & q_{b}
\end{array}\right)=\left(\begin{array}{ccc}
\phantom{-}3 & \phantom{-}2 & \phantom{-}0 \\
-5 & -1 & \phantom{-}0 \\
-4 & -3 & -3
 \end{array}\right).~~
\end{gathered}
\end{align}
The coefficients $c_{u,d}^{ij}$ appearing in Eq.~\eqref{eq:Lag} are determined by performing a $\chi^2$ analysis using the experimentally measured quark masses and CKM matrix elements. In this work, we allow these coefficients to take complex values. The resulting best-fit parameters are presented in Appendix~\ref{app:c_coef}.

The FN charge assignments of the RHNs are especially important in our framework: they control both the light-neutrino mass/mixing structure and leptogenesis. Expanding the last three terms of Eq.~\eqref{eq:Lag} around the FN and electroweak vevs, we obtain
\begin{equation}
\label{eq:Lag_numass}
-\mathcal{L}_{m_{\rm lepton}} = m_{e}^{ij} \overline{e^i_L} e_R^j+ m_{D}^{ij} \overline{\nu^i_L} N_R^j + \frac{1}{2} m_M^{ij} \overline{N^{ci}_{R}} N^j_R + \text{h.c.}\,,  
\end{equation}
where the mass matrices are
\begin{align}
\label{eq:lep_masses}
m_{e}^{ij} =  c_{e}^{ij}  \epsilon^{{n}_e^{ij}} v_{EW} ,~  m_{D}^{ij} =  c_{\nu}^{ij}  \epsilon^{{n}_\nu^{ij}} v_{EW} ,~  m_M^{ij}=c_N^{ij} \epsilon^{n_N^{ij}} M .
\end{align}
The Type-I seesaw mechanism generates the mass matrix for the light neutrinos as
\begin{equation}
\label{eq:numass_Mat}
m_\nu = m_D m_M^{-1} m_D^T\,.
\end{equation}
We adopt the top-down parametrization introduced in Ref.~\cite{Casas:2006hf} to connect high-energy seesaw parameters with low-energy neutrino observables. Without loss of generality, we work in the basis where the RHN Majorana mass matrix is diagonal:
\begin{equation}
m_M = D_{m_M} = \text{diag}(m_{N_1}, m_{N_2}, m_{N_3}),\qquad m_{N_1} \leq m_{N_2} \leq m_{N_3}.
\end{equation}
The Dirac mass matrix in the Lagrangian \eqref{eq:Lag_numass} can be diagonalized via a bi-unitary transformation
\begin{equation}
\label{eq:Diracmass_Mat}
m_D = V_L^* D_{m_D} V_R^{\dagger}\,,
\end{equation}
where $D_{m_D}$= diag $(y_1,y_2,y_3) $, $y_i \geq 0$ and $y_1 \leq y_2 \leq y_3$.
Here \(V_L\) and \(V_R\) are unitary matrices acting in the flavor spaces of the left-handed lepton doublets and RHNs, respectively.
Substituting Eq.~\eqref{eq:Diracmass_Mat} into Eq.~\eqref{eq:numass_Mat}, we obtain
\begin{equation}
\label{eq:mnu}
 m_\nu = m_D D_{m_N}^{-1} m_D^T 
 = V_L^* D_{m_D} V_R^{\dagger} D_{m_N}^{-1} V_R^* D_{m_D} V_L^\dagger\,.
\end{equation}
It is convenient to define
\begin{equation}
 m_\nu' = D_{m_D} V_R^\dagger D_{m_N}^{-1} V_R^* D_{m_D},
\end{equation}
which is independent of the left-handed rotation matrix $V_L$. The light-neutrino mass matrix $m_\nu$ (in Eq.~\eqref{eq:mnu}) can then be diagonalized by first diagonalizing $m_\nu'$ with a unitary matrix $W_L$, such that
\begin{equation}
W_L^T\, m_\nu'\, W_L = D_{m_\nu}.
\end{equation}
The PMNS matrix is therefore given by
\begin{equation}
U_{\rm PMNS} = V_L\, W_L.
\end{equation}

Thus, starting from the high-energy parameters $y_i$, $m_{N_i}$, $V_R$, and $V_L$, one can construct the light-neutrino mass matrix $m_\nu$ in the basis where both the charged-lepton mass matrix $m_e$ and the RHN mass matrix $m_N$ are diagonal.

We emphasize that the DM candidate $N_1$ does not participate in the seesaw mechanism. Consequently, one light neutrino remains massless in this framework. The interactions of $N_1$ arise solely from the last term in Eq.~\eqref{eq:Lag}, yielding the following effective couplings of $N_1$ to the scalar ($s$) and pseudoscalar ($a$) flavon components.
\begin{align}
\label{eq:gDM1}
-\mathcal{L}_\text{DM} \ &\supset \ \frac{1}{2} \, c_N^{11} \epsilon^{n_N^{11}} \left(1 + n_N^{11}\, \frac{s + ia}{\sqrt{2} v_\phi} \right) M \overline{N_{R}^{c\,1}} N_{R}^1 \ + \ \text{h.c.} \nn \\[2ex]
&= \frac{1}{2} \, \mdm \overline{N^1} N^1 + g_{\rm DM} \left( s \overline{N^1} N^1+ i a \overline{N^1} \gamma_5 N^1 \right),
\end{align}
where $m_\text{DM}\equiv m_{N_1}=c_N^{11} \epsilon^{n_N^{11}}M$ is the Majorana mass of the DM candidate $N_1$, and $\gdm = - q_{N_1} \, \mdm/(\sqrt{2} \, v_\phi)$ sets the strength of the scalar ($s$) and pseudoscalar ($a$) portal interactions between the DM sector and SM fermions. This analysis highlights that the FN symmetry-breaking scale $v_\phi$ simultaneously controls flavon-mediated interaction strengths and RHN mass scales, making it central to the model phenomenology.

\section{Constraints from meson mixing}
\label{sec:constraint}
In the FN framework, generation-dependent flavor charges induce non-universal scalar and pseudoscalar couplings to fermions. Consequently, tree-level flavor-changing neutral currents arise from boson exchange. These effects are most pronounced in channels with heavier fermions, since the relevant couplings in Eqs.~\eqref{eq:g+} and \eqref{eq:g-} scale with fermion masses. As shown in Ref.~\cite{Mandal:2023jnv}, direct-search limits mainly from top-quark decays only weakly constrain the viable parameter space. We therefore focus on the most stringent flavor observables, namely neutral-meson mixing constraints relevant to the present analysis.

The off-diagonal element $M_{12}$ in the neutral meson mass matrix
represents $P^0-\overline{P^0}$ mixing as
\begin{equation}
    2m_P (M_{12}^{q*})=\bra{\overline{P^0}}H_{\text{eff}}^{\Delta F=2}\ket {P^0}\,,
\end{equation}
where the factor $2m_P$ reflects the normalization of external states. For $\Delta F=2$, the effective Hamiltonian is given by
 \begin{equation}
 \label{eq:H_mix}
    H^{\Delta F=2}_{\text{eff}}=\dfrac{G^2_F M^2_W}{16\pi^2}\sum C_i(\mu)Q_i(\mu) + {\rm h.c.}\,.
 \end{equation}
The matrix elements of the four-quark operators $Q_i$
sandwiched between the meson states are parametrized in terms of the Bag factor $P_i$ as
\begin{equation}
\bra{\overline{P^0}}Q_i\ket{P^0}=\dfrac{2}{3}F^2_{P_q}m^2_P P_i\,,
\end{equation}
where the $P_i$'s are obtained by renormalization group evolution from the conventional Bag factors $B_i$ computed at the low scale $\mu_b$, namely $P_i = \eta(\mu) B_i$.  The SM contribution is generated dominantly by $Q_1^{VLL}=(\bar{q}\gamma_\mu P_Lq')(\bar{q}\gamma^\mu P_L q')$ four-quark operator and using Eq.~\eqref{eq:H_mix}, we get
\begin{equation}
\label{eq:M12_gen}
    (M^{q*}_{12})_{\rm SM}=\frac{G_F^2 M_W^2}{12\pi^2} F^2_{P_q}m_P P_1^{VLL} \sum_{i,j} V^*_{i,q}V_{i,q'}V^*_{j,q}V_{j,q'} S_0(x_i,x_j)\,,
\end{equation}
where $i,j=u,c,t$ are the internal quarks and $q,q'= b $ or $d$ or $s$. Here
$S_0 (x_i,x_j)$ is the Inami-Lim function~\cite{Inami:1980fz} encoding the loop contributions with $x_{i,j}=m^2_{i,j}/M^2_W$, dealing with the mass of the internal quark(s) entering in the box-diagram and $V_{m,n}$ are the corresponding CKM elements.

\subsection{\texorpdfstring{$B_{d,s}$}{Bds}-meson mixing }
In the case of $B_q$-meson mixing, the dominant SM contribution originates from top-quark loops, which simplifies the expression in Eq.~\eqref{eq:M12_gen}. Including the additional contribution arising from pseudoscalar exchange in our present model, we obtain
\begin{align}
\label{eq:M12_tot}
    M_{12}^{q*}&=\dfrac{G_F^2 M_W^2}{12 \pi^2}P^{VLL}_1 F^2_{B_q}m_{B_q} (V_{t,b}V^*_{t,q})^2 S_0(x_t)
    \nn \\  &+  \dfrac{1}{3}m_BF_B^2P_1^{SLL}(C_1^{SLL}+C_1^{SRR})+\frac{1}{3}m_B F_B^2P_2^{LR}C_2^{LR}\,,
\end{align}
where the Wilson coefficients $C_1^{SLL}$, $C_1^{SRR}$ and $C_2^{LR}$ correspond to the operators induced by the pseudoscalar interaction 
\begin{align}
Q_1^{SLL} =  \left(\bar{b}^\alpha P_L q^\alpha  \right)\left(\bar{b}^\beta  P_L q^\beta  \right)\,, ~
Q_1^{SRR} = \left(\bar{b}^\alpha P_R q^\alpha  \right)\left(\bar{b}^\beta  P_R q^\beta  \right)~{\rm and}~ ~
Q_2^{LR}  =  \left(\bar{b}^\alpha P_L q^\alpha  \right)\left(\bar{b}^\beta  P_R q^\beta  \right)\nn \,,
\end{align}
respectively, are given as
\begin{align}
C_1^{SLL} \ &= \ -\frac{\left[(g_-^d)_{3q}-(g_+^d)_{3q}  \right]^2}{ m_a^2}\,,\nn\\[1ex]
C_1^{SRR} \ &= \ -\frac{\left[(g_-^d)_{3q}+(g_+^d)_{3q}  \right]^2}{ m_a^2}\,,\\[1ex]
C_2^{LR} \ &= \ \frac{\left[(g_-^d)_{3q}+(g_+^d)_{3q}  \right]\left[(g_-^d)_{3q}-(g_+^d)_{3q}  \right]^*}{ m_a^2} \,.\nn
\end{align}
Two key observables directly related to the mixing amplitude $M_{12}^q$ are the neutral meson mass difference,
\begin{equation}
\Delta M_q = 2|M_{12}^q|\,,
\end{equation}
and the time-dependent CP asymmetry measured in decay modes governed by the $b \to sc\bar c$ transition, such as $B \to J/\psi K_S$ and $B_s \to J/\psi \phi$. The latter arises from the interference between $B_q-\bar B_q$ mixing and the decay amplitude to the common final state. It is conventionally expressed as
\begin{equation}
    S_f=\dfrac{2\Im\big[\lambda_f\big]}{1+|\lambda_f|^2}\quad {\rm with}\quad
    \lambda_f=\sqrt{\dfrac{M^{q*}_{12}}{|M^q_{12}|}}\, \dfrac{A(\bar B_q \to J/\psi f )}{A( B_q \to J/\psi f )}\,.
    \label{eq:Sf}
\end{equation}
In the present model, contributions to the $b \to s c\bar c$ transition arise from scalar and pseudoscalar neutral-current operators. As a result, within the factorization approach, these operators do not contribute to the production of a vector charmonium state such as the $J/\psi$ in the final state. Consequently, the corresponding time-dependent CP asymmetry is governed entirely by the phases, both SM and NP, entering the $B_q$-$\bar B_q$ mixing amplitude. The current experimental averages, obtained from a combination of measurements in $B \to J/\psi K_S$, $B \to \psi(2S) K_S$, $B \to \chi_c K_S$ and $B_s \to J/\psi \phi$, $B_s \to \psi(2S) \phi$, provide the respective values of $S_{K_S}$ and $S_{\phi}$ as \cite{HFLAV:2023pne},
\begin{align}
\label{eq:SKS}
S_{K_S} = 0.710 \pm 0.011 \quad {\rm and} \quad
S_\phi = -0.040\pm 0.016\,.
\end{align}
We evaluate the constraints on the model parameter space using the relevant hadronic matrix elements and decay constants. For the Bag parameters, we adopt the values reported in Ref.~\cite{Dowdall:2019bea}. Specifically, for the $B_d$-meson system, relevant to the $B \to J/\psi K_S,~\psi(2S) K_S,~\chi_c K_S$ decay, we use
\begin{align}
    P_1^{VLL} &= 0.673 \pm 0.011\,, \quad 
    P_1^{SLL} = -1.29 \pm 0.13\,, \quad 
    P_2^{LR} = 3.25 \pm 0.32\,.
    \label{eq:Bd_params}
\end{align}
Correspondingly, for the $B_s$-meson system governing $B_s \to J/\psi \phi,~\psi(2S) \phi$, the parameters are
\begin{align}
    P_1^{VLL} &= 0.679 \pm 0.011\,, \quad 
    P_1^{SLL} = -1.30 \pm 0.13\,, \quad 
    P_2^{LR} = 3.25 \pm 0.32\,.
    \label{eq:Bs_params}
\end{align}
For the decay constants, we use the most recent FLAG averages~\cite{Aoki:2024flag}, $f_{B_d}=190.5 \pm 4.3$~MeV and $f_{B_s}=230.7 \pm 3.7$~MeV. Combining these inputs with the expressions for the CP asymmetries $S_{K_S}$ and $S_{\phi}$ derived in Eq.~\eqref{eq:Sf} and the data given in Eq.~\eqref{eq:SKS}, we obtain lower bounds on the product of the pseudoscalar mass $m_a$ and the FN symmetry breaking scale $v_\phi$ as summarized in Table~\ref{tab:constraint}.

\subsection{Kaon mixing}
Analogously, for the kaon system, the mass difference and indirect CP-violation parameter are given by
\begin{equation}
    \Delta M_K=2 \Re[M_{12}^K] \quad {\rm and} \quad \epsilon_K = \frac{e^{i \pi/4}}{\sqrt{2} \Delta M_K} \Im [ M_{12}^K]\,.
\end{equation}
In this case, the SM contribution is dominated by the charm-loop term because of CKM enhancement~\cite{Buchalla:1995vs}:
 \begin{equation}
     (M_{12}^K)_{\rm SM}=\dfrac{G^2_F}{12\pi^2}M_W^2F_K^2 m_K(\lambda_c^2 P_{1c}^{VLL}+\lambda_t^2 P_{1t}^{VLL}+2\lambda_c\lambda_t P_{1ct}^{VLL})\,.
 \end{equation}
Here, $P_{1x}^{VLL}=\eta_x B_K$ includes renormalization-group evolution of the Bag parameter for the corresponding four-quark operator. The experimental values are~\cite{Aoki:2024flag}
\begin{align}
\Delta M_K = (3.484 \pm 0.006) \times 10^{-15}\,\gev, \quad |\epsilon_K| = (2.228 \pm 0.011) \times 10^{-3}\,.
\end{align}
With the appropriate modification of Eq.~\eqref{eq:M12_tot} for the transition from the $B$-meson to the neutral Kaon system ($K^0 - \bar{K}^0$), we incorporate the relevant QCD corrections and hadronic matrix elements. For the SM contributions, we utilize the updated NNLO and NLO estimates for the perturbative factors and the Bag parameter~\cite{Aoki:2024flag,Brod:2010mj}:
\begin{align}
    B_K = 0.773\,, \quad 
    \eta_{cc} = 1.86 \pm 0.53\,, \quad 
    \eta_{tt} = 0.577 \pm 0.007\,, \quad 
    \eta_{ct} = 0.496 \pm 0.047\,.
\end{align}
For the new physics contributions, the combined hadronic factors $P_i \equiv \eta_i B_i$ are determined using the anomalous dimensions from Ref.~\cite{Buras:2000if} to be
\begin{equation}
    P_1^{SLL} = 0.374\,, \quad 
    P_2^{LR} = 3.99\,.
\end{equation}
Using the kaon decay constant $f_K = 155.7$~MeV and the inputs above, we derive a lower bound on the product $m_a v_\phi$. The results are summarized in Table~\ref{tab:constraint}. The bound from $\epsilon_K$ is weaker than those obtained from the $B$-meson observables $S_{K_S}$ and $S_{\phi}$.
\begin{table}[H] 
    \centering
    \renewcommand{\arraystretch}{1.5} 
    \begin{tabular}{|c|c|} 
        \hline
        Observable &  Lower bound in $\gev^2$ \\ \hline \hline
        $ S_{K_S}$           & $ v_\phi m_a \geq  2.14 \times 10^6$    \\ \hline
        $S_\phi$             & $v_\phi m_a \geq  4.18 \times 10^5$    \\ \hline
        $|\epsilon_K|$       & $v_\phi m_a \geq  1 \times 10^5$   \\ \hline \hline
    \end{tabular}
    \caption{
   Summary of lower bounds on $v_\phi\,m_a$ inferred from $B_{d,s}$-meson and kaon mixing observables.}
    \label{tab:constraint}
\end{table}

\section{CP asymmetry and Boltzmann equations for Leptogenesis}
\label{sec:boltzmann}

\begin{figure}[h!]
\centering
\begin{tikzpicture}[baseline=(current bounding box.center)]
  \begin{feynman}
    \vertex (v);
    \vertex [left=1.5cm of v] (in) {\(N_{\beta}\)};
    \vertex [above right=1cm and 1.2cm of v] (out1) {\(H\)};
    \vertex [below right=1cm and 1.2cm of v] (out2) {\(L_\alpha\)};
    \diagram* {
      (in) -- [majorana] (v),
      (v) -- [scalar] (out1),
      (v) -- [fermion] (out2),
    };
    \node at (v) [below left=1pt] {\(\lambda_{\alpha\beta}\)};
  \end{feynman}
\end{tikzpicture}
\hfill
\begin{tikzpicture}[baseline=(current bounding box.center)]
  \begin{feynman}
    \vertex (v1);
    \vertex [left=1.5cm of v1] (in) {\(N_{\beta}\)};
    \vertex [above right=1cm and 1.2cm of v1] (v2);
    \vertex [below right=1cm and 1.2cm of v1] (v3);
    \vertex [right=1.2cm of v2] (out1) {\(H\)};
    \vertex [right=1.2cm of v3] (out2) {\(L_\alpha\)};
    \diagram* {
      (in) -- [majorana] (v1),
      (v1) -- [anti fermion, edge label=\(L_i\)] (v2),
      (v1) -- [scalar, edge label'=\(H\)] (v3),
      (v2) -- [majorana, edge label'=\(N_{j}\)] (v3),
      (v2) -- [scalar] (out1),
      (v3) -- [fermion] (out2),
    };
    \node at (v1) [below left=1pt] {\(\lambda^*_{i\beta}\)};
    \node at (v2) [above=4pt] {\(\lambda_{i j}\)};
    \node at (v3) [below=4pt] {\(\lambda_{\alpha j}\)};
  \end{feynman}
\end{tikzpicture}
\hfill
\begin{tikzpicture}[baseline=(current bounding box.center)]
  \begin{feynman}
    \vertex (v1);
    \vertex [left=1.2cm of v1] (in) {\(N_{\beta}\)};
    \vertex [right=1.2cm of v1] (v2);
    \vertex [right=1.2cm of v2] (v3);
    \vertex [above right=1cm and 1.2cm of v3] (out1) {\(H\)};
    \vertex [below right=1cm and 1.2cm of v3] (out2) {\(L_\alpha\)};
    \diagram* {
      (in) -- [majorana] (v1),
      (v1) -- [scalar, half left] (v2),
      (v1) -- [anti fermion, half right] (v2),
      (v2) -- [majorana, edge label=\(N_{j}\)] (v3),
      (v3) -- [scalar] (out1),
      (v3) -- [fermion] (out2),
    };
    \node at (v1) [below left=2pt] {\(\lambda^*_{i\beta}\)};
    \node at (v2) [below right=1pt] {\(\lambda_{ij}\)};
    \node at ($(v1)!0.5!(v2)+(0,0.6)$)[above] {\(H\)};
    \node at ($(v1)!0.5!(v2)-(0,0.6)$)[below] {\(L_i\)};
    \node at (v3) [right=4pt] {\(\lambda_{\alpha j}\)};
  \end{feynman}
\end{tikzpicture}

\vspace{1.5em} 

\begin{tikzpicture}[baseline=(current bounding box.center)]
    \begin{feynman}
        \vertex (v);
        \vertex [above left=1cm and 1.2cm of v] (in1) {\(N_\beta\)};
        \vertex [below left=1cm and 1.2cm of v] (in2) {\(\phi\)};
        \vertex [above right=1cm and 1.2cm of v] (out1) {\(H\)};
        \vertex [below right=1cm and 1.2cm of v] (out2) {\(L_\alpha\)};
        \diagram* {
            (in1) -- [majorana] (v),
            (in2) -- [scalar] (v),
            (v) -- [scalar] (out1),
            (v) -- [fermion] (out2),
        };
        \node at (v) [left=12pt] {\(\tilde{\lambda}_{\alpha\beta}\)};
    \end{feynman}
\end{tikzpicture}
\hfill
\begin{tikzpicture}[baseline=(current bounding box.center)]
    \begin{feynman}
        \vertex (v1);
        \vertex [above left=1cm and 1.2cm of v1] (in1) {\(N_\beta\)};
        \vertex [below left=1cm and 1.2cm of v1] (in2) {\(\phi\)};
        \vertex [above right=1cm and 1.2cm of v1] (v2);
        \vertex [below right=1cm and 1.2cm of v1] (v3);
        \vertex [right=1.2cm of v2] (out1) {\(H\)};
        \vertex [right=1.2cm of v3] (out2) {\(L_\alpha\)};
        \diagram* {
            (in1) -- [majorana] (v1),
            (in2) -- [scalar] (v1),
            (v1) -- [anti fermion, edge label=\(L_i\)] (v2),
            (v1) -- [scalar, edge label'=\(H\)] (v3),
            (v2) -- [majorana, edge label'=\(N_{j}\)] (v3),
            (v2) -- [scalar] (out1),
            (v3) -- [fermion] (out2),
        };
        \node at (v1) [left=12pt] {\(\tilde{\lambda}^*_{i\beta}\)};
        \node at (v2) [above=4pt] {\(\lambda_{i j}\)};
        \node at (v3) [below=4pt] {\(\lambda_{\alpha j}\)};
    \end{feynman}
\end{tikzpicture}
\hfill
\begin{tikzpicture}[baseline=(current bounding box.center)]
    \begin{feynman}
        \vertex (v1);
        \vertex [above left=1cm and 1.2cm of v1] (in1) {\(N_\beta\)};
        \vertex [below left=1cm and 1.2cm of v1] (in2) {\(\phi\)};
        \vertex [right=1.2cm of v1] (v2);
        \vertex [right=1.2cm of v2] (v3);
        \vertex [above right=1cm and 1.2cm of v3] (out1) {\(H\)};
        \vertex [below right=1cm and 1.2cm of v3] (out2) {\(L_\alpha\)};
        \diagram* {
            (in1) -- [majorana] (v1),
            (in2) -- [scalar] (v1),
            (v1) -- [scalar, half left] (v2),
            (v1) -- [anti fermion, half right] (v2),
            (v2) -- [majorana, edge label=\(N_{j}\)] (v3),
            (v3) -- [scalar] (out1),
            (v3) -- [fermion] (out2),
        };
        \node at (v1) [left=12pt] {\(\tilde{\lambda}^*_{i\beta}\)};
        \node at (v2) [below right=1pt] {\(\lambda_{ij}\)};
        \node at ($(v1)!0.5!(v2)+(0,0.6)$)[above] {\(H\)};
        \node at ($(v1)!0.5!(v2)-(0,0.6)$)[below] {\(L_i\)};
        \node at (v3) [right=4pt] {\(\lambda_{\alpha j}\)};
    \end{feynman}
\end{tikzpicture}

  \caption{Tree-level and one-loop Feynman diagrams driving thermal leptogenesis in the present framework with the standard $1 \to 2$ right-handed neutrino decays ($N_\beta \to L_\alpha H$) and $2 \to 2$ scattering processes involving flavon in the initial state ($N_\beta \phi \to L_\alpha H$).
  }
  \label{fig:dialepto}
\end{figure}

In this section, we present the framework for generating a lepton asymmetry, which is subsequently converted into a baryon asymmetry through the sphaleron processes. We consider thermal leptogenesis in which RHNs are produced through scattering in the thermal bath. A quasi-degenerate $N_2$--$N_3$ spectrum is required to enhance CP violation. The CP asymmetry is sourced by the interference between the tree-level and one-loop amplitudes across two distinct lepton number violating channels: the conventional two body decays of $N_{2,3}$, ($N_{\beta}\rightarrow LH$) and the $2\rightarrow2$ scatterings with the flavon bath, $N_{2,3}\phi\rightarrow LH$\footnote{ The role of analogous scattering channels in the generation of the baryon asymmetry has also been explored in other beyond-the-Standard-Model scenarios; see Refs.~\cite{Bhattacharyaetal2023,Bhattacharyaetal2025,PhysRevD.106.023515}}. All relevant diagrams are shown in Fig.~\ref{fig:dialepto}. In evaluating these amplitudes, we introduced the following shorthand notation for the vertex factors:
\begin{align}
    \lambda_{ij} &= c_{\nu}^{ij} \epsilon^{n^{ij}_{\nu}}\,, \label{eq:couplam} \\
    \tilde{\lambda}_{ij} &= \frac{n^{ij}_\nu}{\sqrt{2}v_\phi} c_{\nu}^{ij} \epsilon^{n^{ij}_{\nu}} 
    = \frac{n^{ij}_\nu}{\sqrt{2}v_\phi} \lambda_{ij} \,. \label{eq:couplama}
\end{align}
Here, $i \in \{e, \mu, \tau\}$ denote the lepton flavors propagating in the loops, while $\beta,j \in \{2,3\}$ (with $\beta \neq j$) index the external and the internal right-handed neutrino, respectively. We can see that $\lambda_{ij}$ is the dimensionless coupling associated with the standard three-point vertex ($N_\beta \to L_\alpha H$), whereas $\tilde{\lambda}_{ij}$ represents the effective coupling for the flavon-induced scattering vertex ($N_\beta \phi \to L_\alpha H$). As is evident from Eq.~\eqref{eq:couplama}, the amplitude for the scattering process is parametrically suppressed by the FN-breaking scale, $v_\phi$, relative to the two body decay.
Let $\mathcal{M}_D$ and $\mathcal{M}_S$ denote the transition amplitudes for the right-handed neutrino decay ($N_\beta \to L_\alpha H$) and the flavon-induced scattering ($N_\beta \phi \to L_\alpha H$), respectively. Each transition amplitude comprises a tree-level contribution along with one-loop vertex and self-energy corrections. Following FN symmetry breaking, the flavon field acquires a vev, and its physical degrees of freedom decompose into a CP-even scalar $s$ and a CP-odd pseudoscalar $a$. Consequently, the flavon scattering proceeds via two distinct channels $\phi \in \{s, a\}$. To evaluate the CP asymmetry, it is convenient to decompose the total matrix element for a given channel $k \in \{D,S\}$ into effective coupling coefficients $c_k$ and reduced kinematic amplitudes $\mathcal{A}_k$:
\begin{align}\mathcal{M}_k &= \mathcal{M}^{\text{tree}}_k+ \left( \mathcal{M}^\text{vert}_k + \mathcal{M}^\text{self}_k \right) \,,\\
\mathcal{M}_k&=c_k\mathcal{A}_k=c^{\text{tree}}_k\mathcal{A}^{\text{tree}}_k+\sum\bigg\{c^{\text{vert}}_k\mathcal{A}^{\text{vert}}_k+c^{\text{self}}_k\mathcal{A}^{\text{self}}_k\bigg\}\,,
\label{eq:Mtotal}\end{align}
which at tree level for decay and scattering are given by 
\begin{align}
    c^{\text{tree}}_D&=-i\lambda_{\alpha\beta}\,,\\
    c^{\text{tree}}_{S,s}&=-i\tilde{\lambda}_{\alpha\beta}\,,\\c^{\text{tree}}_{S,a}&=\tilde{\lambda}_{\alpha\beta}\,,\\
\mathcal{A}_D=\mathcal{A}_S&=\bar{u}_{L_\alpha}(p_2)P_Ru_{N_\beta}(p_1)\,,
\end{align}
respectively. Appendix~\ref{appE} lists the loop amplitudes expressed in Eqs.~\eqref{Eq:loopampli1} and \eqref{Eq:loopampli2}, and the corresponding couplings are defined in Eq.~(\ref{Eq:cverthh}).

To quantify CP violation in RHN decay, we define for each $N_\beta$ the asymmetry parameter
\begin{equation}
\epsilon^D_\beta = \sum_\alpha \frac{\Gamma(N_\beta \longrightarrow L_\alpha H) - \Gamma(N_\beta \longrightarrow \bar{L}_\alpha \bar{H})}{\Gamma(N_\beta \longrightarrow L_\alpha H) + \Gamma(N_\beta \longrightarrow \bar{L}_\alpha \bar{H})}\,,
\label{eq:epsilon}
\end{equation}
and for the scattering case
\begin{equation}
\epsilon^S_\beta = \sum_\alpha \frac{\sigma(N_\beta\phi \longrightarrow L_\alpha H) - \sigma(N_\beta\phi \longrightarrow \bar{L}_\alpha\bar{H})}{\sigma(N_\beta\phi \longrightarrow L_\alpha H) + \sigma(N_\beta\phi \longrightarrow \bar{L}_\alpha\bar{H})}\,,
\label{eq:epsilonScatt}
\end{equation}where we have summed over all flavors.
Normalizing by the total decay rate of each channel keeps the corresponding Boltzmann equations linear in flavor space~\cite{Davidson:2008bu}.

Each CP asymmetry can be written as a sum of interference terms between tree and vertex diagrams ($\epsilon^\text{vert}$), and between tree and self-energy diagrams ($\epsilon^\text{self}$), 
\begin{equation}
\epsilon^{k}_\beta = \epsilon^{\text{vert},\,k}_\beta+ \epsilon^{\text{self},\,k}_\beta \,.
\label{epsilon}
\end{equation}
 The individual contributions, expressed in terms of the reduced amplitudes and phase-space integrals for a generic channel $k$, are given by:
\begin{equation}
    \epsilon^{\text{vert(self)},\,k}_\beta=
    \dfrac{\Im\{c^{\text{tree}}_{k} c^{\text{vert(self)}}_{k}\}}{|c^{\text{tree}}_{k}|^2}\dfrac{2\int \Im\{\mathcal{A}^{\text{tree}}_k\mathcal{A}^{\text{*vert(self)}}_k\}\tilde{\delta}d\Pi_{k}}{\int|\mathcal{A}_k|^2\tilde{\delta}d\Pi_{k}}\,.
\label{eq:eps_defAmp}
\end{equation}
The imaginary part of the amplitude-interference term is obtained using the Cutkosky cutting rules~\cite{Cutkosky:1960sp}, by putting the intermediate particles $L_i$ and $H$ on on-shell and integrating over their phase space:
\begin{equation}
    2\Im{\{\mathcal{A}^{\text{tree}}_k\mathcal{A}^{*\text{vert(self)}}_{k}\}}=\mathcal{A}^{\text{tree}}_k\sum_{i}\int{\mathcal{A}^{*\text{tree}}_{k}(\mathcal{X}_k \rightarrow\bar{L}_i\bar{H})\tilde{\delta'}d\Pi_{i,H}\mathcal{A}^{*\text{tree}}_{k}(\bar{L}_i\,\bar{H}\rightarrow H\,L_\alpha)}\,,
    \label{eq:cutkoskyamp}
\end{equation}
where $\mathcal{X}_k$ represents the initial state(s) for the corresponding channel ($N_\beta$ for the decay and $N_\beta \phi$ for the scattering).
Detailed derivations and amplitude expressions are provided in Appendix~\ref{appE}. Summing over intermediate states $i$ and $j\,(\neq\beta)$, the CP-asymmetry contributions are given in Eqs.~\eqref{eq:std_eps} -- \eqref{eq:eps_hh_vert}. By introducing the normalization factors $N_H = (8\pi[\lambda^*\lambda]_{\alpha\beta})^{-1}$ for decay and $\tilde{N}_H = (8\pi[\tilde\lambda^*\tilde\lambda]_{\alpha\beta})^{-1}$ for scattering, the results can be expressed in terms of the vertex factors (Eqs.~\eqref{eq:couplam} and \eqref{eq:couplama}). The combined vertex and self-energy contributions are:
\begin{align}
\epsilon_{\beta}^{\text{vert},D}+\epsilon_{\beta}^{\text{self},D} &= N_H \sum_{j\neq\beta}\sum_{\alpha,i} \Im\left\{\lambda_{\alpha\beta} \lambda^*_{\alpha j} \lambda^*_{i j} \lambda_{i \beta}\right\}[f(x)+g(x)]\,,\label{eq:std_eps}\\    \epsilon_{\beta}^{\text{vert},S}+\epsilon_{\beta}^{\text{self},S} &= 2z \tilde{N}_H\sum_{j\neq\beta}\sum_{\alpha,i} \Im\{\tilde\lambda_{\alpha\beta}\lambda^*_{\alpha j}\lambda^*_{i j}\tilde\lambda_{i \beta}\}  \nn \\ &\times \dsp\frac{\bigint_{\Psi_\text{min}}^\Lambda d\Psi\sqrt{\lambda(s,M^2_{N_\beta},M^2_\phi)}K_1(\sqrt{\Psi})[g(x_s)-f(x_s)]}{\bigint_{\Psi_\text{min}}^\Lambda d\Psi\sqrt{\dsp\frac{\lambda(s,M^2_{N_\beta},M^2_\phi)}{\Psi}}K_1(\sqrt{\Psi})[\Psi T^2+M^2_{N_\beta}-M^2_{\phi}]}\,,
    \label{eq:eps_hh_vert} 
\end{align}

The mass ratios are given by $x=M^2_{N_j}/M_{N_{\beta}}^2$ and $x_s=M^2_{N_j}/s$, where we neglected the tiny mass of pseudoscalar flavon $a$ in the above expression.  Finally, for $X\in \{x,x_s\}$, the standard Yanagida loop functions~\cite{Fukugita:1986hr} are:
\begin{align}
    f(X)&=\sqrt{X}\Bigg[1-(1+X)\ln\bigg(\dfrac{1+X}{X}\bigg)\Bigg]\,,
    \label{yanagidaform1}\\
    g(X)&=\dfrac{\sqrt{X}}{1-X}\,.
    \label{yanagidaform2}
\end{align}

For the scattering contribution in Eq.~\eqref{eq:eps_hh_vert}, the phase-space integration is performed over the dimensionless variable $\Psi \equiv s/T^2$, where $s$ is the COM energy squared. The integral is bounded from below by the kinematic threshold of the scattering process, $\Psi_{\text{min}} = (M_{N_\beta} + M_\phi)^2/T^2$. Furthermore, because the validity of our effective model is strictly confined to momentum transfers below the FN symmetry-breaking scale, the integration is bounded from above by a hard kinematic cutoff, parameterized by the dimensionless limit $\Lambda$. Finally, $K_1(\sqrt{\Psi})$ denotes the modified Bessel function of the second kind.\\
We now set up the Boltzmann equations governing the evolution of the RHN abundances and lepton asymmetry. Denoting the RHN yields by $Y_{N_\beta}$ and the lepton asymmetry by $Y_L -Y_{\bar L}\equiv Y_{\Delta L}$, the coupled equations in terms of the dimensionless variable $z=M_{N_\beta}/T$ are
\begin{align}
\label{eq:boltz1}
\frac{dY_{N_\beta}}{dz} &= -\bigg [D_\beta(z)+S^s_{\beta}(z)+S^t_{\beta}(z)+S^\phi_\beta(z)\bigg]\left(Y_{N_\beta} - Y_{N_\beta}^{\text{eq}}(z) \right), \\
\label{eq:boltz2}
\frac{dY_{ \Delta L}}{dz} &= \sum_{\beta=2}^{3} \bigg[\left(\epsilon^D_\beta D_\beta(z)+\epsilon^S_\beta S^\phi_\beta(z)\right) \left(Y_{N_\beta} - Y_{N_\beta}^{\text{eq}}(z)\right)
- W^{\rm tot}_\beta(z)  Y_{\Delta L}\bigg].
\end{align}
Note that the evolution of the RHN abundance is governed by decay and scattering processes, while the lepton asymmetry receives contributions from CP-violating decays along with the CP-violating scattering and is damped by washout effects.

Here, the term $D_\beta$ represents the contributions from decay channels, namely $N_{\beta} \to L H $, given by 
\begin{align}
    D_{\beta}=\dfrac{z}{H(M_{N_\beta})}\,\dfrac{K_1(z)}{K_2(z)}\dfrac{M_{N_\beta}}{8\pi}\sum_{\alpha} |\lambda_{\alpha,\beta}|^2 \,,
\end{align}
where the Hubble parameter $H(M_{N_\beta})$ is evaluated at the RHN mass scale and $K_{1,2}(z)$ are the modified Bessel functions.

We include the contributions from $\Delta L = 1$ scattering processes, denoted by $S^{s,t}_\beta(z)$, arising from the $s$-channel process $N_{2,3} L \to t\bar{Q}$ and the $t$-channel process $N_{2,3}Q \to Lt$. We also include the $N_{2,3} \phi \to L H$ scattering contributions as discussed above.
Using superscripts $s$ and $t$ to indicate the corresponding channel contributions for each process, the explicit expressions are given by
\begin{align}
S^s_\beta(z) &= 2|\lambda_{t,q}|^2 \sum_\alpha|\lambda_{\alpha,\beta}|^2K_{\beta}f^s(z), \\
S^t_\beta(z) &= 4|\lambda_{t,q}|^2\sum_\alpha|\lambda_{\alpha,\beta}|^2K_{\beta} f^t(z, y_h),\\
S^\phi_\beta(z) &=\dfrac{1}{4}\sum_\alpha|\tilde\lambda_{\alpha,\beta}|^2 K_\beta f^\phi(z,z_\phi)\,,
\end{align}
with the pre-factor defined as 
\begin{equation}
K_{\beta} = \frac{\, M_\beta}{256 \pi^3 H(M_\beta) }.
\end{equation}

The functions $f^{s,t,\phi}$ are the integrals of the scattering cross-sections, given by
\begin{align}
f^s(z) &= \frac{1}{z^2 K_2(z)} \int_{z^2}^{\infty} d\Psi\, \sigma_s(\Psi/z^2) \sqrt{\Psi} K_1(\sqrt{\Psi}), \\
f^t(z, y_h) &= \frac{1}{z^2 K_2(z)} \int_{z^2}^{\infty} d\Psi\, \sigma_t(\Psi/z^2, y_h) \sqrt{\Psi} K_1(\sqrt{\Psi}),\\
f^\phi(z,z_\phi) &=\dfrac{1}{z^2 K_2(z)}\int^\Lambda_{z^2}d\Psi\,\sigma_\phi(\sqrt{\Psi}/z,\sqrt{\Psi}/z_\phi)\sqrt{\Psi}K_1(\sqrt{\Psi})
\end{align}
where the reduced cross-sections are obtained by incorporating the finite thermal mass effects of the Higgs boson, yielding
\begin{align}
\sigma_s(x) &= \left( \frac{x - 1}{x} \right)^2 , \\
\sigma_t(x, y_h) &= \left( \frac{x - 1}{x} \right) \left[ \frac{x - 2 + 2y_h}{x - 1 + y_h} + \left( \frac{1 - 2y_h}{x - 1} \right) \log\left( \frac{x - 1 + y_h}{y_h} \right) \right],\\
\sigma_\phi(x,x_\phi)&=M^2_{N_\beta}\bigg(x^2-1-\dfrac{x^2}{x^2_\phi}\bigg)\bigg[1-\bigg(\dfrac{1}{x}+\dfrac{1}{x_\phi}\bigg)^2\bigg]^{1/2}
\bigg[1-\bigg(\dfrac{1}{x}-\dfrac{1}{x_\phi}\bigg)^2\bigg]^{1/2}\,.
\end{align}
Here, $y_h$ serves as the infrared regulator for the $t$-channel process. To capture the finite thermal mass effects of the Higgs boson in the early universe bath, we used $m_h(T) \approx 0.4 T$~\cite{Davidson:2008bu} . In terms of the standard evolution variable $z = M_\beta/T$, this yields a dynamic regulator $y_h(z) = 0.16/z^2$. 
Finally, the total washout term in the Boltzmann equations is given by
\begin{equation}
\label{eq:Washout}
W_\beta^{\text{tot}}(z) =\frac{Y_{N_\beta}^{\text{eq}}(z)}{2Y_\ell^{\text{eq}}} \left[ D_\beta(z) + S^s_\beta(z) + S^t_\beta(z) +S^\phi_\beta(z) \right],
\end{equation}
which includes contributions from inverse decays, $\Delta L=1$ scattering processes, and $2\rightarrow2$ scattering processes. We neglect $\Delta L=2$ processes, which arise solely from interactions between RHNs and SM particles, as they are known to give negligible contributions.
The equilibrium yields of the RHNs and the SM leptons entering the Boltzmann equations are given by
\begin{align}
Y_{N_\beta}^{\rm eq} &= \frac{1}{\pi^2 s} T^3 z^2 K_2(z), \\
Y_\ell^{\rm eq} &= \frac{3\zeta(3)}{2\pi^2 s} T^3 ,
\label{Eq:Yleq}
\end{align}
where $s$ denotes the entropy density of the thermal bath. The RHN equilibrium yield follows the Maxwell-Boltzmann distribution appropriate for non-relativistic species, while the lepton equilibrium yield corresponds to relativistic fermions in thermal equilibrium.
Substituting the entropy density,
$s=(g_{*s}2\pi^2T^3)/45$
into Eq.~(\ref{Eq:Yleq}) eliminates the explicit temperature dependence in the equilibrium lepton yield, yielding
\begin{equation}
Y_\ell^{\rm eq} = \frac{135\zeta(3)}{4\pi^4 g_{*s}} \approx 3.9\times10^{-3}.
\end{equation}
Here in the last line, we used Apéry's constant $\zeta(3)\simeq1.202$ and the effective number of relativistic degrees of freedom in the SM, $g_{*s}\simeq106.75$. 
We emphasize that this equilibrium lepton yield provides the appropriate normalization for the washout term in the Boltzmann equations and remains essentially constant throughout the temperature range relevant for leptogenesis, as long as the leptons remain relativistic.

\section{Results} 
\label{sec:results}

We present leptogenesis results for two parameter regions previously identified as compatible with the DM relic abundance in Ref.~\cite{Mandal:2023jnv}. Although the leptogenesis analysis is independent, freeze-in and freeze-out DM production occur in qualitatively different parts of parameter space. These regions correspond to different FN charge assignments, which determine RHN mass spectra, interaction strengths, and seesaw realization. In addition, flavon-mediated effects differ strongly between the two regimes because the FN symmetry-breaking scale $v_\phi$ is directly linked to RHN Majorana masses through Eqs.~\eqref{eq:eps_cabibo} and~\eqref{eq:lep_masses}. We therefore discuss both cases separately.

\subsection{Leptogenesis in freeze-in-compatible parameter space}
\label{sec:freeze-in}

\begin{figure}[H]
 \centering
{\includegraphics[width=.7\linewidth]{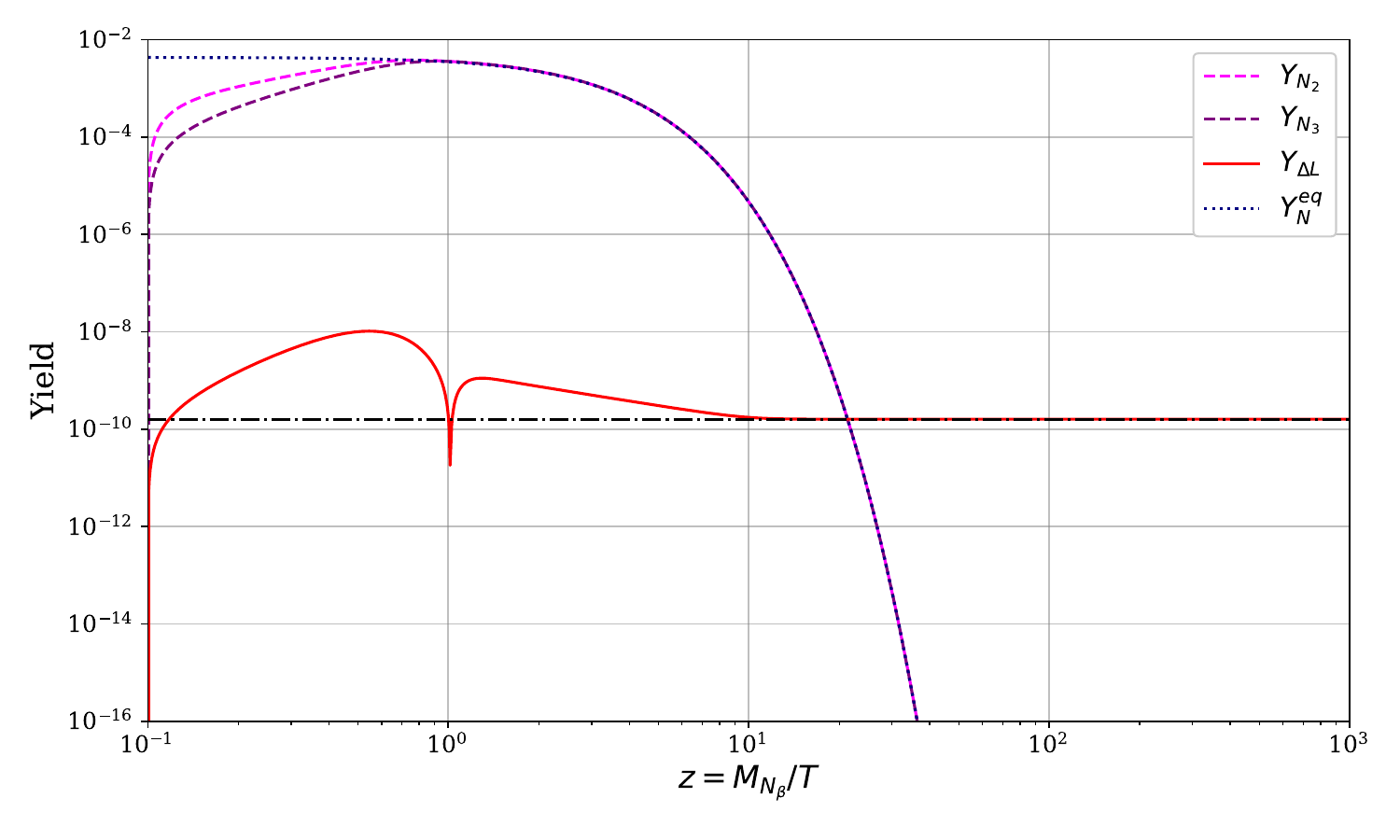}}
\caption{Evolution of the heavy-RHN yields $Y_{N_2}$ (magenta dashed) and $Y_{N_3}$ (purple dashed) as functions of $z=M_\beta/T$ for Benchmark-I in the freeze-in-compatible scenario. The dotted blue curve corresponds to the equilibrium RHN yield, $Y_{N_\beta}^{\rm eq}$, while the solid red curve shows the evolution of the lepton asymmetry, $Y_{\Delta L}$. The generated asymmetry saturates at $Y_{\Delta L} \simeq 1.6\times10^{-10}$, indicated by the black dot-dashed reference line.}
\label{fig:yield}
\end{figure}

In this scenario, the DM relic abundance is generated through freeze-in, with the dominant production channel $aa\to N_1N_1$. Keeping DM out of thermal equilibrium in the early Universe requires a high FN breaking scale, $v_\phi\in[10^7,10^9]$~GeV~\cite{Mandal:2023jnv}. We then investigate whether this region can simultaneously accommodate successful thermal leptogenesis. Note that the lightest RHN, $N_1$, serves as the dark matter candidate and therefore carries an FN charge fixed to $q_{N_1}=-5$ in the freeze-in compatible region. To ensure that $N_1$ remains the lightest RHN, the FN charges of $N_2$ and $N_3$ must satisfy $q_{N_{2,3}} > q_{N_1}$. Consequently, one may choose $q_{N_{2,3}} \in [-4,0]$, allowing the masses of $N_2$ and $N_3$ to span a broad range while preserving the desired mass hierarchy.

\subsubsection{Benchmark-I}
We first focus on the case with the heavier RHNs attainable within the allowed charge assignments, corresponding to $q_{N_{2,3}}=0$. The FN charge assignments used to reproduce the observed lepton masses and mixing pattern are
\begin{align}
\left(\begin{array}{ccc}
q_{L_{1}} & q_{L_{2}} & q_{L_{3}} \\
q_{N_{1}} & q_{N_{2}} & q_{N_{3}} \\
q_{e} & q_{\mu} & q_{\tau}
\end{array}\right) = 
\left(\begin{array}{ccc}
\phantom{-}6 & 5 & 5 \\
-5 & 0 & 0 \\
-3 & 0 & 2
\end{array}\right).
\end{align}
We then perform a $\chi^2$ analysis, varying $c_{e,\nu,N}^{ij}\in[-3,3]$, to refine charged-lepton masses and PMNS parameters. As light-neutrino masses depend on $v_\phi$ (see Eq.~\eqref{eq:numass_Mat}), $v_\phi$ is also treated as a fit parameter to reproduce the observed mass-squared splittings $\Delta m_{21}^2$ and $\Delta m_{32}^2$. We only consider normal ordering. Details are provided in Appendix~\ref{app:Clep_freezein}. The best-fit point gives $\chi^2_{\min}\simeq0.1$, yielding a simultaneous fit to charged-lepton masses, neutrino mass-squared splittings, and PMNS mixing parameters.
At this best-fit point, $v_\phi=3.20\times10^8$~GeV, consistent with successful freeze-in DM production. As $v_\phi$ is large, flavon effects in CP asymmetry are strongly suppressed, and the framework approaches standard thermal leptogenesis. The relevant RHN masses are nearly degenerate, $M_{N_2}=5.89\times10^9$~GeV and $M_{N_3}=5.98\times10^9$~GeV, which enhances CP asymmetries (defined in Eq.~\eqref{eq:epsilon}) to $\epsilon^D_2=5.95\times10^{-6}$ and $\epsilon^D_3=6.19\times10^{-6}$.
Note that the CP-violating scattering process $N_\beta \phi \to H L$ is kinematically inaccessible in this regime. Since the RHN masses satisfy $m_{N_{2,3}}\simeq \mathcal{O}(M)$, i.e. they lie close to the FN cutoff scale, the phase-space integration (in Eq.~\eqref{eq:eps_hh_vert}) is effectively squeezed out, leaving no available range for the upper integration limit $\Lambda$.

Using these benchmark values of $\epsilon^D_\beta$, we solve the coupled Boltzmann equations (Eqs.~\eqref{eq:boltz1} and~\eqref{eq:boltz2} with $S^\phi_\beta(z)=0$), including the $\Delta L=1$ scattering and washout effects. The resulting RHN yields $Y_{N_i}$ and lepton asymmetry $Y_{\Delta L}$ are shown in Fig.~\ref{fig:yield}. The asymmetry initially grows due to CP-violating decays and is subsequently reduced by inverse decays. Once the temperature falls below RHN masses, inverse decays become Boltzmann suppressed ($e^{-M_{N_i}/T}$), and the asymmetry freezes to $Y_{\Delta L}^\infty=1.59\times10^{-10}$. This lepton asymmetry is then converted into a baryon asymmetry through electroweak sphalerons.
Using the standard conversion factor,
$Y_B=28/51 \,Y_{\Delta L}^\infty$~\cite{Davidson:2008bu},
we find
\[Y_B=8.75 \times 10^{-11},\]
in agreement with the cosmic microwave background measurements $Y_B^{\rm obs}=8.75 \times 10^{-11}$ within experimental uncertainty~\cite{Planck:2018vyg}.

\begin{figure}[t]
 \centering
{\includegraphics[width=.6\linewidth]{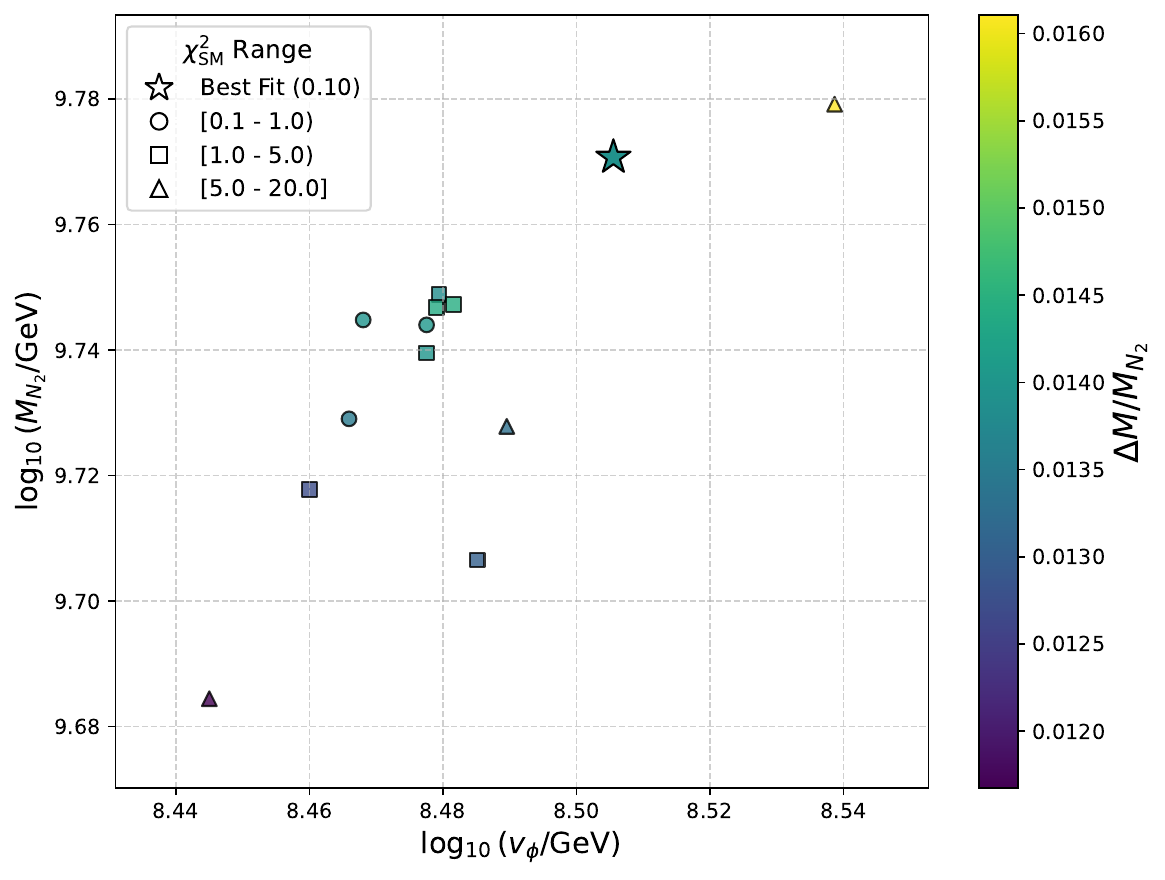}}
\caption{Scan in the $\log v_\phi$--$\log M_{N_2}$ plane in Benchmark-I scenario for points yielding $1.5\times10^{-10} \leq Y_{\Delta L}^{\infty} \leq 1.7\times10^{-10}$. Different marker shapes indicate different $\chi^2$ ranges. The color bar, labeled by $\Delta M/M_{N_2}$, shows the required RHN mass degeneracy.}
\label{fig:scan}
\end{figure}

To map the leptogenesis-compatible region, we scan over $c_{e,\nu,N}^{ij}$ and $v_\phi$, selecting points that yield $1.5\times10^{-10} \leq Y_{\Delta L}^{\infty} \leq 1.7\times10^{-10}$. For each point, we report the associated $\chi^2$ value quantifying agreement with lepton masses, mixing angles, and oscillation data. The result is shown in Fig.~\ref{fig:scan} in the $v_\phi$--$M_{N_2}$ plane. The color bar, $\Delta M/M_{N_2}$ where $\Delta M\equiv |M_{N_2}-M_{N_3}|$, indicates the required RHN mass degeneracy. Thirteen benchmark points are displayed, with marker shapes denoting different $\chi^2$ ranges, demonstrating that successful DM production and leptogenesis can be simultaneously realized over a reasonably broad region of parameter space. The corresponding FN coefficients are listed in Table~\ref{tab:scan_fit} of Appendix~\ref{app:Clep_freezein}.%

\subsubsection{Benchmark-II}

\begin{figure}[H]
 \centering
{\includegraphics[width=.7\linewidth]{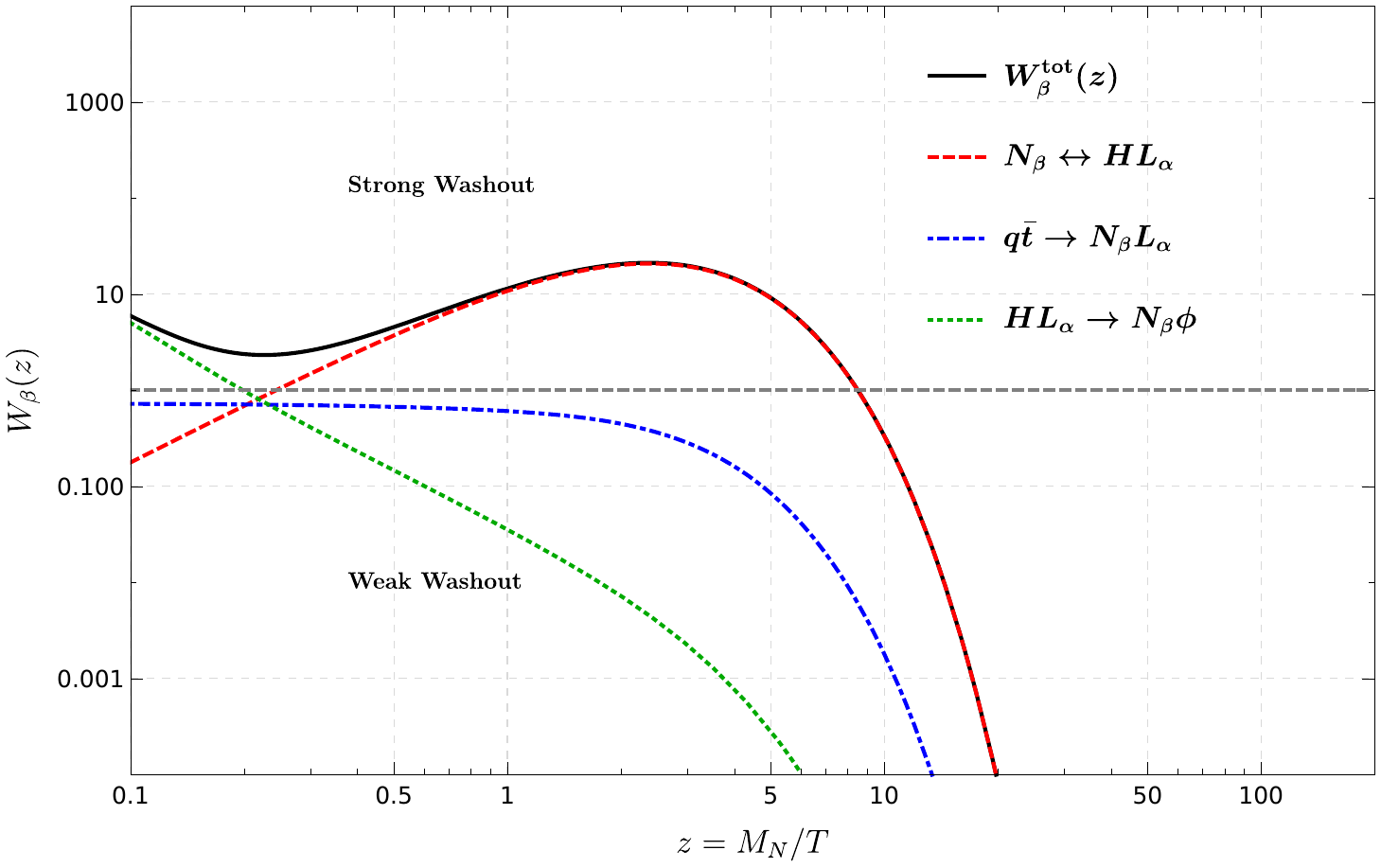}}
\caption{Evolution of the total washout parameter $W^{\text{tot}}_\beta(z)$, defined in Eq.~(\ref{eq:Washout}), as a function of $z=M_N/T$ for Benchmark II. The solid black curve shows the total washout rate, while the colored curves represent the individual contributions. At high temperatures ($z\ll1$), scattering processes contribute significantly to both the washout and the thermal CP asymmetry. As the temperature decreases, inverse decays (red dashed) become dominant and drive the system into the strong-washout regime ($W>1$) around $z\sim1$. For $z\gg1$, all washout processes are Boltzmann suppressed, allowing the generated lepton asymmetry to freeze out.}
\label{fig:washout}
\end{figure}

We next investigate whether lowering the masses of the heavier RHNs opens up regions of parameter space consistent with successful leptogenesis. For this purpose, we scan the allowed range of FN charges $q_{N_{2,3}}$ and determine the charge assignment yielding the lowest $\chi^2$ while remaining consistent with the observed charged-lepton masses, neutrino mass-squared differences, and PMNS mixing parameters. The resulting FN charge assignments are
\begin{align}
\left(\begin{array}{ccc}
q_{L_{1}} & q_{L_{2}} & q_{L_{3}} \\
q_{N_{1}} & q_{N_{2}} & q_{N_{3}} \\
q_{e} & q_{\mu} & q_{\tau}
\end{array}\right) = 
\left(\begin{array}{ccc}
\phantom{-}6 & \phantom{-}5 & \phantom{-}5 \\
-5 & -2 & -2 \\
-3 & \phantom{-}0 & \phantom{-}2
\end{array}\right).
\end{align}
The best-fit result corresponds to $\chi^2_{\min}\simeq1.87$ and $v_\phi=4.22\times10^8$~GeV, consistent with a successful freeze-in DM production region. The heavier RHN masses are nearly degenerate, $M_{N_2}=1.70789\times10^7$~GeV and $M_{N_3}=1.70796\times10^7$~GeV. This benchmark allows RHN-flavor scattering to open up, thereby serving as an additional source of CP violation. In Fig.~\ref{fig:washout}, we compare the contributions of the relevant decay and scattering channels to the total washout rate as a function of $z$. In the relativistic regime ($z \ll 1$), thermal scattering processes contribute significantly to the generation of the lepton asymmetry through temperature-dependent source terms, with the process $N_\beta \phi \to L_\alpha H$ providing the dominant correction (see Eq.~\eqref{eq:epsilonScatt}). As the Universe cools, this channel becomes kinematically suppressed. As a result, the total CP asymmetry smoothly approaches the zero-temperature decay asymmetries, $\epsilon_2^{D} \simeq 4.76 \times 10^{-6}$ and $\epsilon_3^{D} \simeq 7.59 \times 10^{-6}$. Concurrently, the standard $\Delta L=1$ thermal scattering process, $N_\beta L_\alpha \to t \bar Q$, which contributes to the washout at high temperatures, also becomes Boltzmann suppressed, allowing the generated asymmetry to survive.

Solving Boltzmann equations \eqref{eq:boltz1} and~\eqref{eq:boltz2}, including all the contributions from decay, scattering, and washout terms, we find the asymmetry freezes to $Y_{\Delta L}^\infty=1.58\times10^{-10}$ and thus generating the baryon asymmetry 
$Y_B=8.69 \times 10^{-11}$. Note that the slight discrepancy lies well within the theoretical uncertainties associated with the integrated, unflavored Boltzmann treatment. Such uncertainties naturally arise from the neglect of subleading effects, including thermal mass corrections, spectator processes, and momentum-dependent kinetics~\cite{Nardi:2006fx, Giudice:2003jh}.

\begin{figure}[H]
 \centering
{\includegraphics[width=.7\linewidth]{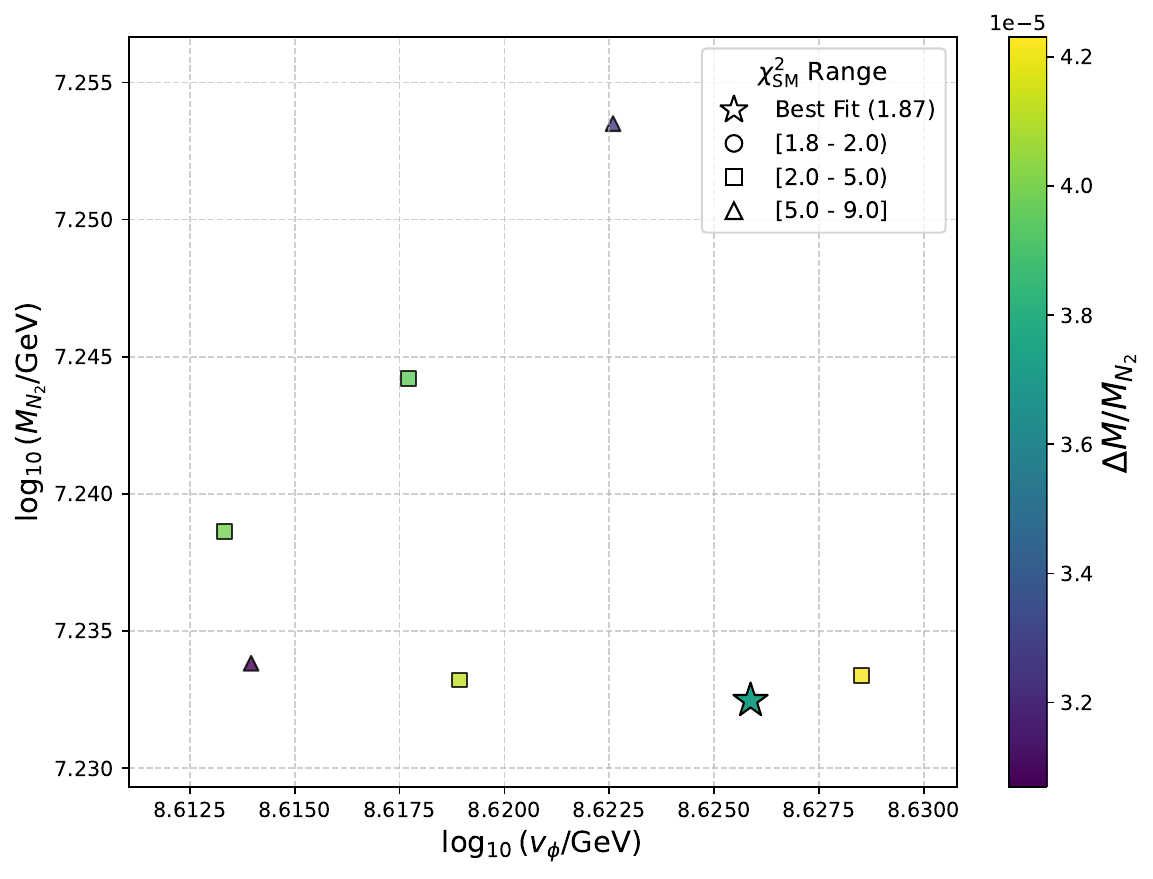}}
\caption{Parameter space scan for the Benchmark-II scenario, showing points that yield a target lepton asymmetry of $1.5\times10^{-10} \leq Y_{\Delta L}^{\infty} \leq 1.7\times10^{-10}$ The corresponding  $\Delta M/M_{N_2}$ color map illustrates that successful leptogenesis in this low-scale regime requires an RHN mass degeneracy approximately three orders of magnitude stronger than in Benchmark-I. }
\label{fig:scanBP2}
\end{figure}

To identify viable leptogenesis solutions in this lower-RHN-mass regime, we perform a numerical scan over the FN coefficients $c_{e,\nu,N}^{ij}$ together with the symmetry-breaking scale $v_\phi$, retaining parameter points that reproduce the observed baryon asymmetry, corresponding to $1.5\times10^{-10} \leq Y_{\Delta L}^{\infty} \leq 1.7\times10^{-10}$. The resulting parameter space is displayed in Fig.~\ref{fig:scanBP2} in the $v_\phi$--$M_{N_2}$ plane, with the associated $\chi^2$ values quantifying the agreement with charged-lepton masses, neutrino oscillation observables, and PMNS mixing parameters. The color coding represents the relative mass splitting $\Delta M/M_{N_2}$, revealing that the required RHN mass degeneracy is approximately three orders of magnitude more severe than in the previous benchmark scenario. This behavior is expected, as thermal leptogenesis in the low-mass RHN regime relies increasingly on resonant enhancement of the CP asymmetry. Nine representative benchmark points are shown, distinguished by marker shapes corresponding to different $\chi^2$ intervals. The associated FN coefficients are provided in Table~\ref{tab:scan_fitBP2} of Appendix~\ref{app:Clep_freezein}.

\subsection{Leptogenesis in freeze-out-compatible parameter space}
\label{sec:freeze-out}
In the freeze-out-compatible scenario, the required FN scale is much lower, $v_\phi\in[1,10]$~TeV, which makes it qualitatively different from the freeze-in case. Since in our setup $M_{N_{2,3}}\simeq v_\phi/\epsilon\simeq5v_\phi$, RHN masses are correspondingly low, disfavoring standard (non-resonant) thermal leptogenesis. Furthermore, maintaining strictly $\mathcal{O}(1)$ Majorana coefficients implies that the RHN masses lie near or above the effective field theory cutoff, $\Lambda_{\rm EFT} \simeq M = v_\phi/\epsilon$. Consequently, $2 \leftrightarrow 2$ flavon-induced scattering processes are kinematically inaccessible within the valid momentum range of the effective theory. The dynamically evolving CP asymmetry is therefore driven purely by the standard $1 \leftrightarrow 2$ thermal decays and inverse decays ($N \leftrightarrow L H$).

Because the CP asymmetry from Eq.~\eqref{eq:epsilon} is too small to generate the required lepton asymmetry in this region, we consider resonant enhancement~\cite{Pilaftsis:2003gt}. In the tiny mass-splitting limit, $|M_{N_3}-M_{N_2}|\ll M_{N_2}$, self-energy contributions dominate through the loop functions $g(x)$ Eq.~\eqref{yanagidaform2}, and the CP asymmetry is
\begin{align}
    \epsilon_\beta = \sum_{j\neq\beta}
    \frac{\text{Im}(\lambda^\dagger \lambda)_{\beta j}^2}{(\lambda^\dagger \lambda)_{\beta\beta}(\lambda^\dagger \lambda)_{jj}} \left[ \frac{(M_{N_\beta}^2 - M_{N_j}^2) M_{N_\beta} \Gamma_{N_j}}{(M_{N_\beta}^2 - M_{N_j}^2)^2 + M_{N_\beta}^2 \Gamma_{N_j}^2} \right] \,,
\end{align}
where $\Gamma_{N_k}$ denotes the decay width of $N_k$. Solving the Boltzmann equations without resonance yields a lepton asymmetry about seven orders of magnitude below the target value. Resonance must therefore enhance $\epsilon_\beta$ from roughly $10^{-13}$ to $10^{-6}$. This requires the $N_2$--$N_3$ mass splitting to satisfy
 \begin{align}
    \Delta M = \left| M_{N_3} - M_{N_2} \right| \sim 5 \times 10^4 \, \Gamma_{N_2} = 5 \times 10^4 \left[ \frac{(\lambda^\dagger \lambda)_{22}}{8\pi} M_{N_2} \right] \sim 10^{-8} \, M_{N_2} \,.
    \label{eq:fine-tuning}
\end{align}

By construction, the FN charge assignments in Eq.~\eqref{eq:FNl_freezein} naturally place $N_2$ and $N_3$ at similar masses. This follows from requiring the DM candidate $N_1$ to be the lightest RHN, with $m_{N_1}\simeq v_\phi$. As shown in Ref.~\cite{Mandal:2023jnv}, this condition strongly constrains viable FN charges and leads to a quasi-degenerate heavier-RHN spectrum.
\begin{align}
\left(\begin{array}{ccc}
q_{L_{1}} & q_{L_{2}} & q_{L_{3}} \\
q_{N_{1}} & q_{N_{2}} & q_{N_{3}} \\
q_{e} & q_{\mu} & q_{\tau}
\end{array}\right) = 
\left(\begin{array}{ccc}
\phantom{-}9 & 8 & 8 \\
-1 & 0 & 0 \\
\phantom{-}0 & 3 & 5
\end{array}\right).
\label{eq:FNl_freezein}
\end{align}

\begin{figure}[t]
 \centering
{\includegraphics[width=.7\linewidth]{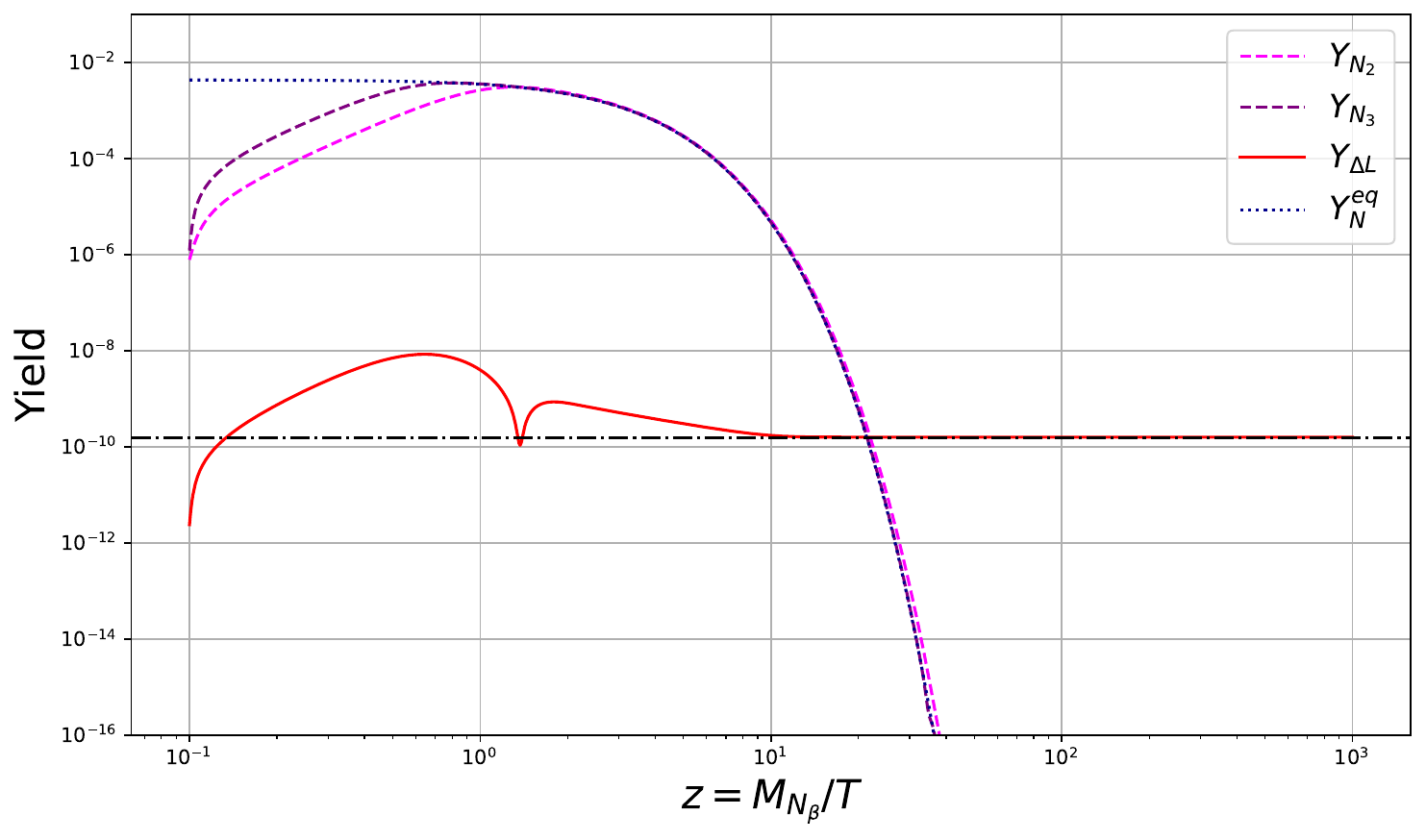}}
\caption{Evolution of RHN yields $N_2$ and $N_3$ as functions of $z=M_{N_\beta}/T$ in the freeze-out-compatible scenario. The color convention is the same as in Fig.~\ref{fig:yield}. }
\label{fig:yield_freeze-out}
\end{figure}

Eq.~\eqref{eq:fine-tuning} indicates that a mass degeneracy at the level of $\mathcal{O}(10^{-7})$ is required between $M_{N_2}$ and $M_{N_3}$ in order to realize resonant leptogenesis. This requirement is explicitly illustrated in the benchmark scenario discussed below, where we solve the Boltzmann equations to determine the final lepton asymmetry. In this case, the RHN masses are highly degenerate, 
with $M_{N_2} \approx 150\,\text{TeV}$ and $\Delta M=10\,\text{MeV}$,
satisfying the fine-tuning condition specified in Eq.~\eqref{eq:fine-tuning}.
This near-degeneracy enhances CP asymmetry through the resonant mechanism, yielding $\epsilon_2 = 8.83\times10^{-6}$ and $\epsilon_3 = 1.68\times10^{-6}$. The corresponding FN scale is $v_\phi=9.6$~TeV. Details of all remaining parameters, including FN coefficients and the predicted $U_{\rm PMNS}$ matrix for this benchmark, are provided in Appendix~\ref{app:Clep_freezeout}. The final lepton asymmetry is $Y_{\Delta L}^\infty = 1.58\times10^{-10}$, corresponding to $Y_B=8.72\times10^{-11}$, consistent with the observed baryon asymmetry. The evolution of lepton asymmetry and RHN yields are shown in Fig.~\ref{fig:yield_freeze-out}.

\section{Summary and discussion}
\label{sec:summary}

In this work, we studied a minimal and unified extension of the Standard Model based on the FN framework, aimed at addressing several open questions in particle physics and cosmology. An abelian $U(1)_{\rm FN}$ flavor symmetry, spontaneously broken by a complex flavon field, generates observed hierarchies in fermion masses and mixings through higher-dimensional operators. Extending the setup with three RHNs enables a simultaneous explanation of neutrino masses (via Type-I seesaw), DM, and baryogenesis.

Allowing complex effective FN coefficients is central to this construction. These couplings reproduce the observed CP-violating phases in both quark and lepton sectors, yielding a consistent description of CKM and PMNS matrices. They also induce nontrivial flavor signatures, especially in the quark sector, leading to testable effects in observables such as neutral-meson mixing and CP-violating $B$ decays.

Building on our earlier work, where the lightest RHN $N_1$ provides the correct DM relic abundance through freeze-in or freeze-out, we focused here on the origin of baryon asymmetry. In this analysis, $N_1$ remains effectively decoupled from leptogenesis, while $N_2$ and $N_3$ generate lepton asymmetry through out-of-equilibrium CP-violating decays. We included both the standard decay channels of the RHNs and additional flavon-induced scattering channels in our analysis. These effects introduce additional CP-violating sources and modify decay and/or scattering rates of heavier RHNs. The resulting lepton asymmetry is then converted to baryon asymmetry through electroweak sphalerons.

A distinctive feature of this framework is its restrictive structure: all relevant mass scales and interaction strengths in RHN and flavon sectors are controlled by a single parameter, the FN symmetry-breaking scale $v_\phi$. This differs from many leptogenesis models where RHN masses and Yukawa couplings are largely independent. Here, the same FN charge assignments that make DM viable also tightly constrain the RHN spectrum, yielding a highly correlated setup.

We have demonstrated that successful thermal leptogenesis can be achieved in regions of parameter space compatible with both freeze-in and freeze-out DM scenarios. In the freeze-in regime, characterized by a large symmetry-breaking scale $v_\phi \sim \mathcal{O}(10^7\!-\!10^9),\mathrm{GeV}$, we considered two benchmark cases: one in which the heavy RHN masses lie close to the cutoff scale of the effective theory, and another in which they remain well below it. In both cases, the flavon-mediated scattering processes are strongly suppressed by the large value of $v_\phi$, causing the dynamics to approach those of conventional thermal leptogenesis.

In contrast, the freeze-out-compatible region corresponds to a much lower symmetry-breaking scale, $v_\phi \sim \mathcal{O}(1\!-\!10),\mathrm{TeV}$. In this regime, the RHN masses are pushed towards the cutoff scale, and the resulting kinematic suppression effectively eliminates the $2\leftrightarrow2$ flavon-mediated scatterings. As a consequence, the generation of the lepton asymmetry becomes predominantly decay-driven. Successful leptogenesis then requires a resonant enhancement of the CP asymmetry, implying a significant degree of RHN mass degeneracy.

In conclusion, this work establishes the FN framework as a coherent and economical effective theory that simultaneously addresses flavor hierarchies, neutrino masses, CP violation, DM, and baryogenesis. A natural next step is to investigate an ultraviolet-complete realization, which could clarify the origin of FN symmetry and its dynamics while opening broader phenomenological opportunities across low- and high-energy searches.

\subsubsection*{Acknowledgments}
R.M. acknowledges support from the DAE-BRNS YSRP grant No. 57/20/02/2024 and SERB/ANRF Grant SPG/2022/001238. R.M. also thanks the Deutsche Forschungsgemeinschaft (DFG, German Research Foundation) for support under Grant No. 396021762 (TRR 257) through a Mercator Fellowship during the visit to Universit\"at Siegen. K.R. acknowledges financial support from the University Grants Commission (UGC), Government of India, under the Junior Research Fellowship (JRF) scheme (NTA Ref. No. 231610069466).

\appendix
\section{Determination of Froggatt-Nielsen coefficients from \texorpdfstring{$\chi^2$}{chi2} analysis}
\label{app:c_coef}

\begin{table}[ht]
\centering
\begin{tabular}{|c|c|}
\hline
Parameter & Value \\
\hline
\multicolumn{2}{|c|}{CKM Parameters \cite{CKMfitter2025}} \\
\hline
$\lambda$ & $0.22504 \pm 0.00022$ \\
\hline
$A$ & $0.8215 \pm 0.0146$ \\
\hline
$\bar{\rho}$ & $0.1562 \pm 0.0102$ \\
\hline
$\bar{\eta}$ & $0.3564 \pm 0.0065$ \\
\hline
\multicolumn{2}{|c|}{PMNS Parameters \cite{NuFIT2025}} \\
\hline
$\theta_{12}$ (deg) & $33.68 \pm 0.73$ \\
\hline
$\theta_{23}$ (deg) & $43.3 \pm 0.9$ \\
\hline
$\theta_{13}$ (deg) & $8.56 \pm 0.11$ \\
\hline
$\delta_{CP}$ (deg) & $212 \pm 41$ \\
\hline
$\Delta m_{21}^2\,(10^{-5}\,\mathrm{eV}^2)$ & $7.49 \pm 0.19$ \\
\hline
$\Delta m_{32}^2\,(10^{-3}\,\mathrm{eV}^2)$ & $2.513 \pm 0.021$ \\
\hline
\end{tabular}
\caption{Input values of the CKM and PMNS parameters used in the numerical analysis taken from latest global fit results. }
\label{tab:input CKM and PMNS parameters}
\end{table}

This appendix presents the results of the numerical $\chi^2$ analysis performed to determine the values of the FN coefficients $c_x^{ij}$ appearing in the Lagrangian in Eq.~\eqref{eq:Lag}. We first discuss the results obtained for the quark sector, followed by the lepton sector, where we present results corresponding to parameter regions compatible with both freeze-in and freeze-out scenarios. Our primary objective is to reproduce the experimentally measured fermion masses and mixing parameters up to the three decimal places by varying the $\mathcal{O}(1)$ coefficients $c_x^{ij}$. 

Since in this analysis the FN coefficients are allowed to take complex values, the number of free parameters exceeds the number of experimental observables, which include the quark and lepton masses as well as the elements of the CKM and PMNS matrices. Consequently, the model is under-determined, implying that the solution is not unique. In principle, infinitely many parameter combinations can reproduce the experimental data within their uncertainties. However, the requirement that all FN coefficients remain of $\mathcal{O}(1)$ significantly restricts the viable parameter space, resulting in only a limited number of acceptable solutions that simultaneously yield accurate fermion masses and mixing patterns. Hence, the best-fit solution obtained by minimizing the $\chi^2$ function constructed from these observables should not be interpreted as a global minimum of the $\chi^2$ function. Rather, it should be regarded as a relative measure for comparing different viable solutions within the restricted parameter space. 
We use the Particle Data Group values for all quark and charged-lepton masses. The CKM matrix elements are determined using the current averages of the Wolfenstein parameters reported by Ckmfitter global fit results \cite{CKMfitter2025}, while the PMNS matrix is constructed from the latest NuFIT results \cite{NuFIT2025}, combining IceCube-24 with Super-Kamiokande atmospheric data for normal ordering. All numerical inputs employed in the analysis are summarized in Table~\ref{tab:input CKM and PMNS parameters}.

\subsection{Fit to quark masses and CKM parameters}
\label{app:Cquark}

We quote one set of optimized coefficient matrices for the up-type $(c_u^{ij})$ and down-type $(c_d^{ij})$ quark sectors as provided below.
\begin{equation}
    c_u =
\begin{pmatrix}
-2.26 + 0.22\,i & \phantom{-}0.03 + 0.08\,i & -0.28 + 0.46\,i \\
\phantom{-}0.99 - 1.63\,i & \phantom{-}0.09 + 0.98\,i & \phantom{-}0.49 + 1.40\,i \\
-0.52 - 1.05\,i & -0.04 - 0.25\,i & -0.23 - 0.96\,i
\end{pmatrix}\,,
\end{equation}

\begin{equation}
    c_d =
\begin{pmatrix}
\phantom{-}0.82 - 0.03\,i & \phantom{-}0.33 - 0.75\,i & -0.73 + 0.48\,i \\
\phantom{-}0.15 - 0.24\,i & -1.01 - 0.30\,i & \phantom{-}0.48 + 1.26\,i \\
-0.02 - 1.90\,i & \phantom{-}1.54 + 0.67\,i & \phantom{-}0.12 - 0.93\,i
\end{pmatrix}\,.
\end{equation}
The resulting predictions for the quark masses are
\begin{equation}
\begin{aligned}
(m_u, m_c, m_t) &= (2.16~\text{MeV},\; 1.27~\text{GeV},\; 172.95~\text{GeV}) \,,\\
(m_d, m_s, m_b) &= (4.70~\text{MeV},\; 93.50~\text{MeV},\; 4.18~\text{GeV}) \,,
\end{aligned}
\end{equation}
and the corresponding CKM matrix is
\begin{equation}
    V_{\text{CKM}} =
\begin{pmatrix}
\phantom{-}0.9747  & \phantom{-}0.2234  & \phantom{-}0.00138 - 0.003331 i \\
-0.2233  & \phantom{-}0.9739 & \phantom{-}0.0407  \\
\phantom{-}0.00775 - 0.003331i & -0.0399 & 0.9992
\end{pmatrix}\,,
\end{equation}
which are in good agreement with data.
This optimization yielded a minimum $\chi^2$ value of $\chi^2_{\rm min}=44$, with the dominating contribution originating from the $V_{cs}$ element of the CKM matrix, which alone contributes $\sim 28$ to the total $\chi^2_{\rm min}$. Note that $V_{cs}$ depends only on the Wolfenstein parameter $\lambda$, which is determined with very small uncertainty from global fits (see Table~\ref{tab:input CKM and PMNS parameters}); consequently, the theoretical uncertainty in $V_{cs}$ is correspondingly reduced.

\subsection{Fit to lepton masses and PMNS parameters in freeze-in-compatible region}
\label{app:Clep_freezein}

In addition to the FN coefficients $c_{e,\nu,N}^{ij}$, the FN symmetry-breaking scale $v_\phi$ remains a free parameter in predicting the charged lepton masses and the leptonic mixing matrix, as the light-neutrino masses depend explicitly on it. As discussed in Sec.~\ref{sec:freeze-in}, compatibility with the freeze-in DM scenario requires $v_\phi$ to be sufficiently large in order to keep the DM candidate out of thermal equilibrium. We find that with Benchmark-I, the numerical optimization yields a minimum value of $\chi^2_{\rm min} \approx 0.1$ at a FN symmetry-breaking scale of $v_\phi = 3.20 \times 10^{8}\,\text{GeV}$. The optimized coefficients for the charged lepton $(c_e^{ij})$, light-neutrino $(c_\nu^{ij})$ and RHN $(c_N^{ij})$ sectors are found to be:
\begin{subequations}
\label{eq:benchmark_matrices}
\begin{align}
    c_e &=
    \begin{pmatrix}
\phantom{-}0.36 - 2.89\, i & 0.86 - 0.82\, i & -0.18 - 0.50\, i \\
-0.70 - 1.21\, i & 0.42 - 0.07\, i & -0.18 - 0.42\, i \\
\phantom{-}2.23 - 0.13\, i & 1.23 + 0.11\, i & \phantom{-}0.65 + 0.22\, i \\
\end{pmatrix} , \\[2ex]
    c_{\nu} &=
    \begin{pmatrix}
0 & -1.41 + 3.36\, i & -1.07 + 2.99\, i \\
0 & \phantom{-}2.27 - 3.00\, i & -2.31 + 0.79\, i \\
0 & -0.46 + 1.34\, i & \phantom{-}2.79 - 1.25\, i \\
\end{pmatrix} , \\[2ex]
   c_N&=\begin{pmatrix}
3.53 & 0.00 & 0.00 \\
0.00 & 4.24 & 0.00 \\
0.00 & 0.00 & 4.29 \\
\end{pmatrix}\,.
\end{align}
\end{subequations}

Similarly for the Benchmark-II case (with FN charge assignment $q_{N_{2,3}}=-2$), the numerical optimization yields a minimum value of $\chi^2_{\rm min} \approx 1.87$ at an FN symmetry-breaking scale of $v_\phi = 4.23 \times 10^{8}\,\text{GeV}$. The optimized coefficients for the charged lepton ($c_e$), light-neutrino ($c_\nu$), and RHN ($c_N$) sectors read:

\begin{subequations}
\label{eq:benchmark_matrices}
\begin{align}
    c_e &=
    \begin{pmatrix}
    -0.54 - 2.89i & -1.67 + 2.83i & -0.46 - 0.54i \\
    \phantom{-}0.26 - 0.46i & -0.56 - 0.34i & \phantom{-}0.04 - 0.01i \\
    -0.55 - 0.43i & -0.59 + 0.73i & -0.78 + 0.27i
    \end{pmatrix} , \\[2ex]
    c_{\nu} &=
    \begin{pmatrix}
    0 & -1.00 + 1.89i & -3.02 + 2.91i \\
    0 & \phantom{-}3.60 + 1.35i & -0.24 - 2.35i \\
    0 & \phantom{-}0.72 - 2.23i & -0.92 + 2.35i
    \end{pmatrix} , \\
\intertext{and}
    c_N &=
    \begin{pmatrix}
    2.38 & 0 & 0 \\
    0 & 3.32 & 0 \\
    0 & 0 & 3.32
    \end{pmatrix} .
\end{align}
\end{subequations}
The corresponding PMNS matrix for both the benchmarks are given by
\begin{align}
    U_{\rm PMNS}=&
\begin{pmatrix}
\phantom{-}0.82 - 0.03\, i & 0.55  & -0.12 + 0.08\, i \\
-0.33 + 0.05\, i & \phantom{-}0.65 + 0.03\, i & 0.68  \\
\phantom{-}0.46 + 0.03\, i & -0.52 + 0.03\, i & 0.72  \\
\end{pmatrix}, {\rm Benchmark-I}\,, \\
    U_{\rm PMNS}=&
\begin{pmatrix}
 \phantom{-}0.8259 - 0.0159\,i &  0.5494 & -0.1163 + 0.0473\,i \\
-0.3383 + 0.0346\,i &  \phantom{-}0.6528 + 0.0161\,i &  0.6767 \\
 \phantom{-}0.4490 + 0.0190\,i & -0.5209 + 0.0208\,i &  0.7254 
\end{pmatrix}, {\rm Benchmark-II}\,,
\end{align}
The resulting predictions for the charged-lepton masses, light-neutrino mass-squared differences, and RHN masses are presented in the the first two columns of Table~\ref{tab:lepton_comparison}. 

\begin{table}[ht]
\centering
\renewcommand{\arraystretch}{1.2} 
\begin{tabular}{|c|c|c|c|}
\hline
Parameters & Freeze-in (BP-I) & Freeze-in (BP-II) & Freeze-out  \\
and masses & $q_N=(0,0)$ & $q_N=(-2,-2)$ & compatible region \\ \hline
$\chi^2_{\rm min}$ & 0.10 & 1.87 & 1.5 \\ \hline

$v_\phi$ [GeV]  & $3.20 \times 10^8$ & $4.23 \times 10^8$ & $9651.4$ \\ \hline

\multicolumn{4}{|c|}{Charged lepton masses} \\ \hline
$m_e$ [MeV] & $0.511$ & $0.511$ & $0.511$ \\ \hline
$m_\mu$ [MeV] & $105.7$ & $105.7$ & $105.7$ \\ \hline
$m_\tau$ [GeV] & $1.78$ & $1.77$ & $1.78$ \\ \hline

\multicolumn{4}{|c|}{Light neutrino mass-squared splittings [$\text{eV}^2$]} \\ \hline
$\Delta m^2_{12}$ & $7.49 \times 10^{-5}$ & $7.51 \times 10^{-5}$ & $7.49 \times 10^{-5}$ \\ \hline
$\Delta m^2_{13}$  & $2.51 \times 10^{-3}$ & $2.51 \times 10^{-3}$ & $2.51 \times 10^{-3}$ \\ \hline

\multicolumn{4}{|c|}{Right-handed neutrino masses [GeV]} \\ \hline
$M_{N_1}$ & $2036.5$ & $3432.1$ & $4802.7$ \\ \hline
$M_{N_2}$ & $5.90 \times 10^9$ & $1.71 \times 10^7$ & $1.54 \times 10^5$ \\ \hline
$M_{N_3}$ & $5.98 \times 10^9$ & $1.71 \times 10^7$ & $1.54 \times 10^5$ \\ \hline
\end{tabular}
\caption{Summary of the $\chi^2$-analysis results along with the predictions for lepton masses in the freeze-in (Benchmarks I and II) and freeze-out-compatible scenarios.}
\label{tab:lepton_comparison}
\end{table}

\begin{table}[p]
\centering
\begin{sideways}
\scriptsize
\renewcommand{\arraystretch}{1.2}
\begin{adjustbox}{max width=0.95\textheight}
\begin{tabular}{ |c|c|c|c|c| }
\hline
\textbf{$\chi^2$} & \textbf{$v_{\phi}$} & \textbf{$C_e$} & \textbf{$C_\nu$} & \textbf{$C_N$} \\
\hline
0.10 &
$3.20\times10^{8}$ &
$\begin{pmatrix}
0.3561-2.8908i & 0.8589-0.8177i & -0.1839-0.5018i \\
-0.6969-1.2057i & 0.4216-0.0667i & -0.1845-0.4236i \\
2.2259-0.1333i & 1.2349+0.1104i & 0.6532+0.2162i \\
\end{pmatrix}$ &
$\begin{pmatrix}
0 & -1.4140+3.3619i & -1.0654+2.9939i \\
0 & 2.2688-3.0027i & -2.3110+0.7900i \\
0 & -0.4621+1.3421i & 2.7936-1.2532i \\
\end{pmatrix}$ &
$\begin{pmatrix}
3.5302 & 0 & 0 \\
0 & 4.2356 & 0 \\
0 & 0 & 4.2944 \\
\end{pmatrix}$ \\
\hline
0.56 &
$3.00\times10^{8}$ &
$\begin{pmatrix}
0.5472-2.6874i & 0.9841-0.2045i & -0.1651-0.4982i \\
-0.7669-1.4522i & 0.5566-0.0776i & -0.1238-0.4208i \\
1.2749-1.8209i & 1.0138-0.1195i & 0.6603+0.2399i \\
\end{pmatrix}$ &
$\begin{pmatrix}
0 & -0.9039+3.2949i & -1.0214+2.7529i \\
0 & 2.2839-3.3010i & -2.3867+0.4351i \\
0 & -0.3779+1.5588i & 2.6496-1.0895i \\
\end{pmatrix}$ &
$\begin{pmatrix}
3.1900 & 0 & 0 \\
0 & 4.2481 & 0 \\
0 & 0 & 4.3077 \\
\end{pmatrix}$ \\
\hline
0.67 &
$2.92\times10^{8}$ &
$\begin{pmatrix}
0.6404-2.9071i & 0.9313-0.3776i & 0.0313-0.6097i \\
-0.8726-1.1643i & 0.6088-0.4136i & -0.1410-0.4142i \\
0.8782-1.0203i & 1.3224+0.1757i & 0.6595+0.2369i \\
\end{pmatrix}$ &
$\begin{pmatrix}
0 & -1.1205+3.2322i & -1.0167+2.6765i \\
0 & 2.1435-3.3714i & -2.3475+0.4532i \\
0 & -0.3551+1.5604i & 2.5778-1.1041i \\
\end{pmatrix}$ &
$\begin{pmatrix}
3.5290 & 0 & 0 \\
0 & 4.2154 & 0 \\
0 & 0 & 4.2722 \\
\end{pmatrix}$ \\
\hline
0.87 &
$2.94\times10^{8}$ &
$\begin{pmatrix}
1.2974-2.6874i & 0.6731+0.0175i & -0.0304-0.6378i \\
-0.7669-1.4522i & 0.5869-0.2488i & -0.1431-0.4462i \\
0.3655-1.8209i & 1.1339-0.0460i & 0.6472+0.2072i \\
\end{pmatrix}$ &
$\begin{pmatrix}
0 & -0.9039+3.2949i & -1.0214+2.7460i \\
0 & 2.2839-3.3108i & -2.3867+0.4351i \\
0 & -0.3717+1.5415i & 2.6345-1.1212i \\
\end{pmatrix}$ &
$\begin{pmatrix}
3.3006 & 0 & 0 \\
0 & 4.3499 & 0 \\
0 & 0 & 4.4106 \\
\end{pmatrix}$ \\
\hline
1.14 &
$3.01\times10^{8}$ &
$\begin{pmatrix}
0.7937-1.9299i & 0.9350+0.1684i & -0.2094-0.5518i \\
-0.8614-1.3084i & 0.6714-0.0035i & -0.1539-0.4162i \\
-0.4964-1.2191i & 0.8444+0.1311i & 0.6436+0.2742i \\
\end{pmatrix}$ &
$\begin{pmatrix}
0 & -0.8328+3.2911i & -0.9606+2.7504i \\
0 & 2.2693-3.3598i & -2.4520+0.4126i \\
0 & -0.3926+1.5563i & 2.6683-1.0829i \\
\end{pmatrix}$ &
$\begin{pmatrix}
3.5172 & 0 & 0 \\
0 & 4.2608 & 0 \\
0 & 0 & 4.3221 \\
\end{pmatrix}$ \\
\hline
1.29 &
$3.03\times10^{8}$ &
$\begin{pmatrix}
0.3366-1.7862i & 1.1104-0.3439i & -0.2011-0.5587i \\
-0.8594-1.6775i & 0.8190-0.0510i & -0.1084-0.4382i \\
0.0072+2.0150i & 0.9442+0.6212i & 0.6466+0.2380i \\
\end{pmatrix}$ &
$\begin{pmatrix}
0 & -0.9652+3.2929i & -0.9741+2.7804i \\
0 & 2.3093-3.3341i & -2.4494+0.3953i \\
0 & -0.3832+1.5423i & 2.6666-1.0703i \\
\end{pmatrix}$ &
$\begin{pmatrix}
4.0238 & 0 & 0 \\
0 & 4.2407 & 0 \\
0 & 0 & 4.3018 \\
\end{pmatrix}$ \\
\hline
1.60 &
$3.00\times10^{8}$ &
$\begin{pmatrix}
0.1239-2.4413i & 1.0400+0.0286i & -0.1715-0.6291i \\
-1.1688-1.3650i & 0.8812-0.0235i & -0.0815-0.4342i \\
0.1629+0.8067i & 0.8576+1.1765i & 0.6378+0.2643i \\
\end{pmatrix}$ &
$\begin{pmatrix}
0 & -1.0304+3.2847i & -0.9506+2.7394i \\
0 & 2.2350-3.3581i & -2.4659+0.4304i \\
0 & -0.3738+1.5322i & 2.6556-1.0232i \\
\end{pmatrix}$ &
$\begin{pmatrix}
3.9345 & 0 & 0 \\
0 & 4.2041 & 0 \\
0 & 0 & 4.2629 \\
\end{pmatrix}$ \\
\hline
1.81 &
$3.06\times10^{8}$ &
$\begin{pmatrix}
-0.1298-2.4109i & 1.2874-0.6585i & -0.2595-0.8714i \\
-1.0852-1.1154i & 1.4070-0.3809i & -0.0945-0.3059i \\
0.9234+1.0796i & 0.8598+2.2653i & 0.6794+0.2717i \\
\end{pmatrix}$ &
$\begin{pmatrix}
0 & -0.9646+3.0154i & -0.9788+2.7328i \\
0 & 1.6726-3.5411i & -2.4078+0.6107i \\
0 & -0.3929+1.4668i & 2.5176-1.1322i \\
\end{pmatrix}$ &
$\begin{pmatrix}
3.0302 & 0 & 0 \\
0 & 3.8296 & 0 \\
0 & 0 & 3.8795 \\
\end{pmatrix}$ \\
\hline
2.11 &
$3.02\times10^{8}$ &
$\begin{pmatrix}
0.0607-2.2744i & 1.0287-0.6297i & -0.1543-0.8369i \\
-0.7600-1.4855i & 0.6849-0.3188i & -0.0330-0.4546i \\
1.8730+3.2566i & 1.2355-0.0636i & 0.6573+0.1481i \\
\end{pmatrix}$ &
$\begin{pmatrix}
0 & -0.8971+3.3826i & -0.9177+2.7771i \\
0 & 2.3021-3.3341i & -2.4411+0.4654i \\
0 & -0.2977+1.5551i & 2.6142-1.1541i \\
\end{pmatrix}$ &
$\begin{pmatrix}
3.0965 & 0 & 0 \\
0 & 4.2774 & 0 \\
0 & 0 & 4.3364 \\
\end{pmatrix}$ \\
\hline
3.94 &
$2.88\times10^{8}$ &
$\begin{pmatrix}
0.0964-2.8633i & 0.9554-0.7727i & 0.0007-0.7688i \\
-0.7576-1.2887i & 1.1531-1.0610i & 0.0246-0.4318i \\
1.6925+2.4351i & 1.3966+0.2185i & 0.6880+0.0639i \\
\end{pmatrix}$ &
$\begin{pmatrix}
0 & -1.3663+3.1595i & -0.9368+2.7480i \\
0 & 2.1040-3.3047i & -2.3696+0.4429i \\
0 & -0.2442+1.5706i & 2.5029-1.2668i \\
\end{pmatrix}$ &
$\begin{pmatrix}
3.0835 & 0 & 0 \\
0 & 4.1644 & 0 \\
0 & 0 & 4.2173 \\
\end{pmatrix}$ \\
\hline
6.47 &
$2.79\times10^{8}$ &
$\begin{pmatrix}
0.1908-2.0018i & 1.7007-0.1421i & -0.0762-0.4805i \\
-0.7250-0.9139i & 1.1308-1.4206i & -0.1433-0.3517i \\
0.9991+2.3505i & 2.6223+2.0629i & 0.6985+0.1413i \\
\end{pmatrix}$ &
$\begin{pmatrix}
0 & -1.6924+2.9254i & -0.7993+2.7139i \\
0 & 1.6754-3.6339i & -2.6718+0.4711i \\
0 & -0.2200+1.4562i & 2.4007-1.1664i \\
\end{pmatrix}$ &
$\begin{pmatrix}
2.0258 & 0 & 0 \\
0 & 3.9918 & 0 \\
0 & 0 & 4.0384 \\
\end{pmatrix}$ \\
\hline
7.64 &
$3.09\times10^{8}$ &
$\begin{pmatrix}
0.5970-2.3545i & 1.6436-0.6536i & -0.1245-0.6954i \\
-0.4522-1.7837i & 1.3212-1.3838i & -0.0719-0.3609i \\
0.8869+1.5305i & 1.7035+1.7143i & 0.6689+0.2603i \\
\end{pmatrix}$ &
$\begin{pmatrix}
0 & -1.9413+3.0224i & -1.0408+2.8421i \\
0 & 2.0045-3.4954i & -2.5512+0.4960i \\
0 & -0.2534+1.5178i & 2.5436-1.2201i \\
\end{pmatrix}$ &
$\begin{pmatrix}
2.4148 & 0 & 0 \\
0 & 3.9811 & 0 \\
0 & 0 & 4.0340 \\
\end{pmatrix}$ \\
\hline
16.44 &
$3.46\times10^{8}$ &
$\begin{pmatrix}
0.0015-1.2872i & 1.4195-0.5874i & -0.2689-1.2514i \\
0.4803-2.0599i & 0.3011-0.1312i & 0.0164-0.4907i \\
0.1929-3.4226i & -1.1505+1.2145i & 0.6218+0.0483i \\
\end{pmatrix}$ &
$\begin{pmatrix}
0 & -1.8455+3.2095i & -1.0579+2.9220i \\
0 & 2.1733-3.6583i & -2.6298+0.5137i \\
0 & -0.1537+1.5935i & 2.5608-1.5499i \\
\end{pmatrix}$ &
$\begin{pmatrix}
1.8478 & 0 & 0 \\
0 & 4.0017 & 0 \\
0 & 0 & 4.0662 \\
\end{pmatrix}$ \\
\hline
\end{tabular}
\end{adjustbox}
\end{sideways}
\caption{Results for $\chi^2$-analysis and FN coefficients for 13 benchmark points shown in Fig.~\ref{fig:scan} in the freeze-in-compatible scenario Benchmark-I.}
\label{tab:scan_fit}

\end{table}

\begin{table}[p]
\centering
\begin{sideways}
\scriptsize
\renewcommand{\arraystretch}{1.2}
\begin{adjustbox}{max width=0.95\textheight}
\begin{tabular}{ |c|c|c|c|c| }
\hline
\textbf{$\chi^2$} & \textbf{$v_{\phi}$} & \textbf{$C_e$} & \textbf{$C_\nu$} & \textbf{$C_N$} \\
\hline
1.87 &
$4.23\times10^{8}$ &
$\begin{pmatrix}
-0.5408-2.8945i & -1.6672+2.8307i & -0.4619-0.5353i \\
0.2609-0.4608i & -0.5619-0.3447i & 0.0413-0.0064i \\
-0.5489-0.4298i & -0.5875+0.7336i & -0.7755+0.2668i \\
\end{pmatrix}$ &
$\begin{pmatrix}
0 & -0.9992+1.8946i & -3.0169+2.9083i \\
0 & 3.5981+1.3544i & -0.2391-2.3469i \\
0 & 0.7193-2.2321i & -0.9195+2.3544i \\
\end{pmatrix}$ &
$\begin{pmatrix}
2.3775 & 0 & 0 \\
0 & 3.3221 & 0 \\
0 & 0 & 3.3222 \\
\end{pmatrix}$ \\
\hline
1.96 &
$4.23\times10^{8}$ &
$\begin{pmatrix}
-0.5408-2.8945i & -1.6672+2.8307i & -0.4619-0.5353i \\
0.2609-0.4608i & -0.5619-0.3447i & 0.0413-0.0064i \\
-0.5489-0.4298i & -0.5875+0.7336i & -0.7755+0.2668i \\
\end{pmatrix}$ &
$\begin{pmatrix}
0 & -0.9992+1.8946i & -3.0169+2.9083i \\
0 & 3.5981+1.3544i & -0.2391-2.3469i \\
0 & 0.7193-2.2321i & -0.9195+2.3544i \\
\end{pmatrix}$ &
$\begin{pmatrix}
2.3775 & 0 & 0 \\
0 & 3.3221 & 0 \\
0 & 0 & 3.3222 \\
\end{pmatrix}$ \\
\hline
2.52 &
$4.25\times10^{8}$ &
$\begin{pmatrix}
-0.8489-2.4794i & -2.5177+2.9353i & -0.3925-0.5523i \\
0.1831-0.4220i & -0.3337-0.4313i & 0.0248-0.0358i \\
-0.0219-0.1310i & 0.0838+0.6415i & -0.7775+0.2665i \\
\end{pmatrix}$ &
$\begin{pmatrix}
0 & -1.0710+1.8245i & -3.1005+2.9085i \\
0 & 3.5178+1.5033i & -0.1431-2.3413i \\
0 & 0.7612-2.2422i & -0.8949+2.3530i \\
\end{pmatrix}$ &
$\begin{pmatrix}
1.0836 & 0 & 0 \\
0 & 3.3089 & 0 \\
0 & 0 & 3.3090 \\
\end{pmatrix}$ \\
\hline
2.87 &
$4.16\times10^{8}$ &
$\begin{pmatrix}
-0.9392-1.6613i & -1.8729+3.5510i & -0.6280-0.4379i \\
-0.0786-0.4884i & -0.4498-0.4233i & -0.0140-0.0062i \\
-0.5975-0.9025i & -0.2802+1.6612i & -0.7610+0.2927i \\
\end{pmatrix}$ &
$\begin{pmatrix}
0 & -1.0825+1.8000i & -3.0945+2.9860i \\
0 & 3.4656+1.2895i & -0.3576-2.3474i \\
0 & 0.8581-2.2004i & -0.8018+2.3849i \\
\end{pmatrix}$ &
$\begin{pmatrix}
1.0956 & 0 & 0 \\
0 & 3.3814 & 0 \\
0 & 0 & 3.3815 \\
\end{pmatrix}$ \\
\hline
3.84 &
$4.15\times10^{8}$ &
$\begin{pmatrix}
-0.9416-1.8996i & -0.9121+3.5353i & -0.6580-0.0337i \\
-0.3178-0.5110i & -0.6183-0.5529i & -0.0409-0.0051i \\
-1.4746-0.9667i & -0.0187+2.1102i & -0.7716+0.2719i \\
\end{pmatrix}$ &
$\begin{pmatrix}
0 & -1.1831+1.9066i & -2.9944+3.2446i \\
0 & 3.4719+1.0894i & -0.5931-2.2587i \\
0 & 0.8961-2.1910i & -0.7615+2.5216i \\
\end{pmatrix}$ &
$\begin{pmatrix}
1.5768 & 0 & 0 \\
0 & 3.4778 & 0 \\
0 & 0 & 3.4780 \\
\end{pmatrix}$ \\
\hline
4.93 &
$4.11\times10^{8}$ &
$\begin{pmatrix}
-1.3512-2.4252i & -1.5489+1.8356i & -0.5239+0.0658i \\
-0.0559-0.3870i & -0.7389-0.4556i & -0.0180+0.0018i \\
-0.4994-0.6945i & 0.1064+1.8256i & -0.7797+0.2562i \\
\end{pmatrix}$ &
$\begin{pmatrix}
0 & -1.1624+1.9589i & -3.0931+3.1116i \\
0 & 3.4702+1.1720i & -0.4971-2.2820i \\
0 & 0.7561-2.2207i & -0.7298+2.5181i \\
\end{pmatrix}$ &
$\begin{pmatrix}
1.8968 & 0 & 0 \\
0 & 3.4683 & 0 \\
0 & 0 & 3.4684 \\
\end{pmatrix}$ \\
\hline
6.24 &
$4.19\times10^{8}$ &
$\begin{pmatrix}
-0.0419-2.3352i & -1.2350+2.9857i & -0.3388-0.0839i \\
0.0645-0.2011i & -0.3683-0.3796i & 0.0392+0.0107i \\
-0.7992-1.0784i & -0.6057-1.0524i & -0.7951+0.2438i \\
\end{pmatrix}$ &
$\begin{pmatrix}
0 & -1.3595+1.6543i & -2.9980+3.5556i \\
0 & 3.4803+1.1034i & -0.6809-2.1802i \\
0 & 0.8800-2.1493i & -0.6656+2.6042i \\
\end{pmatrix}$ &
$\begin{pmatrix}
1.6812 & 0 & 0 \\
0 & 3.5134 & 0 \\
0 & 0 & 3.5135 \\
\end{pmatrix}$ \\
\hline
8.71 &
$4.11\times10^{8}$ &
$\begin{pmatrix}
-1.6560-1.5083i & -2.3923+1.9042i & -0.4421-0.2179i \\
-0.1706-0.5346i & -1.0355-0.2072i & -0.0729-0.0469i \\
0.7251-1.3503i & -0.8972+1.6777i & -0.7928+0.1978i \\
\end{pmatrix}$ &
$\begin{pmatrix}
0 & -1.3027+1.6355i & -3.0354+3.2426i \\
0 & 3.4615+1.3711i & -0.3979-2.3475i \\
0 & 0.5299-2.1899i & -0.6032+2.5326i \\
\end{pmatrix}$ &
$\begin{pmatrix}
1.8848 & 0 & 0 \\
0 & 3.4251 & 0 \\
0 & 0 & 3.4252 \\
\end{pmatrix}$ \\
\hline
35.96 &
$4.03\times10^{8}$ &
$\begin{pmatrix}
-0.3012+0.3527i & -1.5789+1.6053i & -0.4216-0.6690i \\
-0.8132-0.8037i & -0.3573-0.6417i & -0.0689-0.0197i \\
1.0347-0.6317i & 0.4369+2.2476i & -0.8034+0.0497i \\
\end{pmatrix}$ &
$\begin{pmatrix}
0 & -1.5167+1.6439i & -2.8757+3.9077i \\
0 & 3.3306+1.1372i & -0.7342-1.9482i \\
0 & 0.7246-2.1642i & -0.4399+2.8085i \\
\end{pmatrix}$ &
$\begin{pmatrix}
1.9724 & 0 & 0 \\
0 & 3.6716 & 0 \\
0 & 0 & 3.6717 \\
\end{pmatrix}$ \\
\hline
\end{tabular}
\end{adjustbox}
\end{sideways}
\caption{Results for $\chi^2$-analysis and FN coefficients for 9 benchmark points shown in Fig.~\ref{fig:scanBP2} in the freeze-in-compatible scenario Benchmark-II.}
\label{tab:scan_fitBP2}
\end{table}

\newpage
\subsection{Fit to lepton masses and PMNS parameters in freeze-out-compatible region}
\label{app:Clep_freezeout}

For the freeze-out-compatible scenario (see Sec.~\ref{sec:freeze-out}), we find the optimization reaches a minimum value of $\chi^2_{\rm min} = 1.5$ at a FN symmetry-breaking scale of $v_\phi = 9651.4$ GeV.  
The following are the matrices representing the coefficients for the charged lepton $(c_e^{ij})$, light-neutrino $(c_\nu^{ij})$ and RHN $(c_N^{ij})$ sectors, respectively.

\begin{subequations}
\label{eq:flavor_matrices_sub}
\begin{align}
    c_e^{ij} &=
    \begin{pmatrix}
    \phantom{-}0.40 - 1.89i & -0.93 - 0.12i & \phantom{-}0.58 + 0.05i \\
    -0.11 + 0.56i & \phantom{-}0.69 + 0.21i & -0.69 - 0.45i \\
    -0.46 + 0.02i & -0.81 - 0.43i & -0.09 + 0.02i
    \end{pmatrix} , \\[2ex]
    c_{\nu}^{i \alpha} &=
    \begin{pmatrix}
    0 & 0.13 + 2.14i & \phantom{-}1.43 + 0.28i \\
    0 & 0.02 - 0.56i & \phantom{-}0.61 + 1.12i \\
    0 & 0.06 - 0.50i & -1.46 - 0.63i
    \end{pmatrix} , \\
\intertext{and}
    c_{N}^{\alpha \beta} &=
    \begin{pmatrix}
    2.16357066 & 0 & 0 \\
    0 & 3.67119276 & 0 \\
    0 & 0 & 3.671193004
    \end{pmatrix} .
\end{align}
\end{subequations}

The resulting predictions for charged-lepton masses, light-neutrino mass-squared differences, and RHN masses are listed in the third column of Table~\ref{tab:lepton_comparison}. As discussed in Sec.~\ref{sec:freeze-out}, a fine-tuning at the level of $\mathcal{O}(10^{7})$ appears in $c_N^{22}$ and $c_N^{33}$, generating the quasi-degenerate $N_{2,3}$ spectrum required for resonant leptogenesis.
The PMNS matrix in this case is given by,
\begin{equation}
U_{\rm PMNS}=
\begin{pmatrix}
\phantom{-}0.8233 + 0.0169\, i & 0.5487 & -0.1024 + 0.1015 \,i \\
-0.3488 + 0.0536 \,i & \phantom{-}0.6476 + 0.0337 \,i & 0.6745 \\
\phantom{-}0.4391 + 0.0679\, i & -0.5257 + 0.0456 \,i & 0.7240 
\end{pmatrix}\,,
\end{equation}
demonstrating good agreement with current data.

\section{One-loop amplitudes for CP asymmetry calculation}
\label{appE}
In this appendix, we briefly outline the calculation of the CP asymmetry discussed in Sec.~\ref{sec:boltzmann}. We begin by presenting the explicit expressions for the one-loop amplitudes introduced in Eq.~\eqref{eq:Mtotal}, corresponding to the decay and the $2\leftrightarrow2$ flavon scattering processes depicted in Fig.~\ref{fig:dialepto}, given by
\begin{align}
    \mathcal{A}^{\text{vert}} &= \int \frac{d^4 q}{(2\pi)^4} \,
    \frac{
    \bar{u}_{\ell_\alpha}(p_2)\, P_R\left( \slashed{q} - \slashed{p}_3 + M_{N_j} \right)P_R \left( \slashed{q} + M_{L_i} \right)P_L\,u_{N_\beta}(p_1)
    }{
    \left( q^2 - M_{L_i}^2 \right)
    \left( (q - p_3)^2 - M_{N_j}^2 \right)
    \left( (q-p_3 - p_2 )^2 - M_H^2 \right)
    }\,, \label{Eq:loopampli1} \\
    \mathcal{A}^{\text{self}} &= \int \frac{d^4 q}{(2\pi)^4} \,
    \frac{
    \bar{u}_{\ell_\alpha}(p_2)\, P_R\left( \slashed{p}_2 + \slashed{p}_3 + M_{N_j} \right)P_R \left( \slashed{q} + M_{L_i} \right)P_L\,u_{N_\beta}(p_1)
    }{
    \left( q^2 - M_{L_i}^2 \right)
    \left( (q-p_3 - p_2 )^2 - M_H^2 \right)
    }\,.
    \label{Eq:loopampli2}
\end{align}
Here, $p_1$ denotes the momentum of the incoming RHN, while $p_2$ and $p_3$ correspond to the momenta of the outgoing lepton and scalar, respectively. The loop momentum $q$ is carried by the internal fermion $L_i$. For the flavon scattering, $N_\beta(p_1)+\phi(p_k)\rightarrow L_\alpha(p_2)+H(p_3)$, we would evaluate these amplitude in the centre of mass frame with $s=(p_1+p_k)^2=(p_2+p_3)^2$. The corresponding couplings are given by,
\begin{equation}
\begin{aligned}
    c^{\text{vert}}_D = c^{\text{self}}_D &= \lambda_{\alpha j} \lambda_{i j} \lambda^*_{i \beta}
    \label{Eq:cverthh}\,,\\
    c^{\text{vert}}_S = c^{\text{self}}_S &= \lambda_{\alpha j} \lambda_{i j} \tilde\lambda^*_{i \beta}\,.
\end{aligned}
\end{equation}
 
To evaluate the absorptive part of the one-loop amplitude, we employ the Cutkosky cutting rules, which relate the imaginary part of a loop diagram to a phase-space integral over on-shell intermediate states. As an illustrative example, we consider the contribution to the CP asymmetry, defined in Eq.~\eqref{eq:cutkoskyamp}, arising from the interference between the tree-level amplitude and the one-loop vertex diagram for the $N_\beta \to \nu_\alpha H$ channel. We first compute the corresponding imaginary part:
\begin{align}
    2 \Im \left\{ \mathcal{A}^{\text{tree}} \, \mathcal{A}^{\text{*vert}} \right\} &= \mathcal{A}^{\text{tree}} \int d\Pi_{i,H}\,\tilde{\delta'} \, \frac{\Tr \left\{ (\slashed{p}_2 + M_{L_\alpha}) P_R (\slashed{p}_1 + M_{N_\beta}) P_R \slashed{q} P_L M_{N_j} \right\}}{(q - p_3)^2 - M_{N_j}^2} \,,
    \label{eq:internalphasespace}
\end{align}
where $d\Pi_{i,H}\tilde{\delta'}$  represents the measure for the on-shell intermediate state, defined as
\begin{equation}
    d\Pi_{i,H}\,\tilde{\delta'} \equiv \dfrac{d^3 \vec{p}_{{L_i}}}{(2\pi)^3 2E_{L_i}}\dfrac{d^3 \vec{p}_{H}}{(2\pi)^3 2E_H}(2\pi)^4\delta^{(4)}(p_1-p_{L_i}-p_H) \,.
\end{equation}
Working in the rest frame of the decaying RHN, where $p_1 = (M_{N_\beta}, \vec{0})$, and taking the massless limit for all particles except the RHNs, the on-shell conditions for the internal states imply
$|\Vec{p}_{L_i}|=|\vec{p}_H|$ and $E_{L_i}=E_H=M_{N_\beta}/2\,.$ Using these kinematic relations, we perform the Dirac traces and evaluate the phase-space integral over the cut diagram, which yields
   \begin{align}\label{intH}
         2 \Im{\{A^{\text{tree}}_D\,A^{\text{*vert}}_D\}}=\dfrac{M^2_{N_\beta} M_{N_j}}{16\pi}\int d\cos\theta\dfrac{E_2-|\vec{p}_2|\cos\theta}{p^2_2-M_{N_\beta}(E_2+|\vec{p}_2|\cos\theta)-M^2_{N_j}}\,,
    \end{align}
where $\theta$ denotes the angle between the momenta of the internal light neutrino and the outgoing lepton. The numerator of the CP asymmetry, defined in Eq.~\eqref{eq:eps_defAmp}, is obtained by integrating the interference term over the final-state phase space. The corresponding measure for the external particles, $d\Pi_{L, H}\tilde{\delta}$, is defined as
\begin{equation}
    d\Pi_{L_\alpha, H}\tilde{\delta} \equiv \frac{d^3 \vec{p}_{L_\alpha}}{(2\pi)^3 2E_{L_\alpha}} \frac{d^3 \vec{p}_{H}}{(2\pi)^3 2E_H} (2\pi)^4 \delta^{(4)}(p_{N_\beta} - p_{L_\alpha} - p_H) \,.
    \label{eq:ExternalPhaseSpace}
\end{equation}
Performing the integration for the vertex correction yields
\begin{equation}\label{2}
    2\int  d\Pi_{L,H}\Tilde{\delta}\, \Im{\{A^{\text{tree}}_D\,A^{\text{*vert}}_D\}} = \dfrac{M^2_{N_\beta}}{64\pi^2}\sqrt{x}\left(1-(1+x)\ln{\left(\dfrac{1+x}{x}\right)}\right)\,,
\end{equation}
where $x \equiv M^2_{N_j}/M^2_{N_\beta}$ is the squared mass ratio of the heavy neutrinos.
We repeat analogous calculations for the self-energy diagram. Evaluating these contributions, we obtain
\begin{align}\label{3}
    2\int  d\Pi_{L,H}\Tilde{\delta}\, \Im{\{A^{\text{tree}}_D\,A^{\text{*self}}_D\}}& = \dfrac{M^2_{N_\beta}}{64\pi^2}\left(\dfrac{\sqrt{x}}{1-x}\right)\,.
\end{align}
We now turn to the denominator of Eq.~\eqref{eq:cutkoskyamp}, which is obtained by evaluating the tree-level two-body decay width, given by
\begin{align}
\int  d\Pi_{L, H}\Tilde{\delta} |A^{\text{tree}}_D|^2 
= \frac{M^2_{N_\beta}}{8\pi}\,.
\label{eq:cutkoskyamplidenom}
\end{align}
This completes the derivation of the contribution to the CP asymmetry arising from the decay process $N_\beta \to L_\alpha H$.\\ 
We can perform a similar calculation for the flavon scattering phase space. Starting from Eq.~(\ref{eq:internalphasespace}), the two-body phase space measure $d\Pi_{i,H}\tilde{\delta}'$ evaluated in the center-of-mass (COM) frame becomes
\begin{equation}
    d\Pi_{i,H}\,\tilde{\delta}' \equiv \frac{d^3 \vec{p}_{L_i}}{(2\pi)^3 2E_{L_i}} \frac{d^3 \vec{p}_H}{(2\pi)^3 2E_H} (2\pi)^4 \delta^{(4)}(P - p_{L_i} - p_H) \,,
\end{equation}
where $P = (\sqrt{s}, \vec{0})$ is the total four-momentum of the system. In this frame, the kinematics of the incoming state ($N_\beta, \phi$) and the effectively massless outgoing state ($L_i, H$) are governed by the following relations:
\begin{align}
    E_\beta &= \frac{s + M^2_{N_\beta} - M^2_\phi}{2\sqrt{s}}\,, \\
    E_\phi &= \frac{s - M^2_{N_\beta} + M^2_\phi}{2\sqrt{s}}\,, \\
    |\vec{p}_1| &= \frac{\sqrt{\lambda(s, M^2_{N_\beta}, M^2_\phi)}}{2\sqrt{s}}\,, \\
    E_{L_i} &= E_H = |\vec{p}_3| = \frac{\sqrt{s}}{2}\,.\label{eq:COMkine}
\end{align}
We can then perform similar calculations as in the decay case, with appropriate changes in the kinematics in the center of mass frame. The total invariant mass squared of the system is now $s$ rather than $M_{N_\beta}^2$, meaning the energies of the on-shell internal loop states become $\sqrt{s}/2$ as shown in Eq. (\ref{eq:COMkine}). When performing the angular integration over the intermediate momentum, this kinematic shift effectively replaces the mass ratio argument $x = M_{N_j}^2/M_{N_\beta}^2$ in the loop functions with the variable $x_s \equiv M_{N_j}^2/s$. The overall spinor trace evaluates similarly, preserving the $M_{N_\beta} M_{N_j}$ prefactor. Applying these kinematic substitutions, we obtain the following results for the vertex and self-energy diagrams,
\begin{align}
     2\int d\Pi_{L,H} \Tilde{\delta}\, \Im{\{A^{\text{tree}}_S\,A^{\text{*vert}}_S\}} &= \dfrac{M_{N_\beta}M_{N_j}}{64\pi^2}\left(1-(1+x_s)\ln{\left(\dfrac{1+x_s}{x_s}\right)}\right)\,,\\
      2\int  d\Pi_{L,H}\Tilde{\delta}\, \Im{\{A^{\text{tree}}_S\,A^{\text{*self}}_S\}} &= \dfrac{M_{N_\beta}M_{N_j}}{64\pi^2}\left(\dfrac{1}{1-x_s}\right)\,.
\end{align}
As for the denominator we have
\begin{equation}
    \int d\Pi_{L, H} \Tilde{\delta} |A^{\text{tree}}_S|^2 
= \frac{1}{16\pi}(s+M^2_{N_\beta}-M^2_\phi)\,.
\label{eq:cutkoskyamplidenomscatter}
\end{equation}
\bibliographystyle{bibstyle}
\bibliography{biblio.bib}

\end{document}